Adoption of Sustainable Agricultural Practices among

Kentucky Farmers and Their Perception about Farm

Sustainability

Bijesh Mishra

Master Thesis

Submitted to

The Graduate Faculty

of

Kentucky State University

Frankfort, Kentucky

In the Partial Fulfillment of the Requirement for the

Degree of

Master of Science in Environmental Studies

May, 2017

**THESIS ABSTRACT**

**Adoption of Sustainable Agricultural Practices among Kentucky Farmers and Their Perception about Farm Sustainability**

**Bijesh Mishra**

Sustainability of agriculture and food system has been a concern for scientists for long time. The adoption of sustainable agricultural practices (SAPs) has been very helpful to attain agricultural sustainability. However, practices are localized and site specific. The purpose of this thesis was to identify commonly adopted SAPs and their adoption among Kentucky farmers. The specific objectives were: 1) to explore farmer's Perceptions about Farm and Farming Practice Sustainability, 2) to identify predictors of SAPs adoption using farm attributes, farmers' attitudes and behaviors, socioeconomic and demographic factors, and knowledge, and 3) to evaluate adoption barriers of SAPs among Kentucky Farmers. One thousand surveys were sent to farmers using double stratified sampling method throughout

Kentucky in 2015 and 230 responses were analyzed using Negative Binomial Regression.

Kentucky famers operate average of 169.57 acreage land. The average age of Kentucky farmers is 62.85 years. Most of them have formal education below college degree. However, farmers generally perceive that their farm and farming activities attain the objectives of sustainable agriculture. The results of negative binomial regression analysis suggest that row crop growers *(p = 0.000)*, farmers in favor of diversification *(p = 0.000)*, and formal education level *(p = 0.000)* increase probability of SAPs adoption. Vegetable growers *(p = 0.049)*, and farmers with irrigation facilities *(p = 0.026)* also increase probability of the adoption. Income from agro-tourism *(p = 0.074)*, land operated *(p = 0.057)*, age *(p = 0.057)* are other significant variables of the adoption.

Inadequate knowledge, perceived difficulty of implementation, lack of (adequate) market, negative attitude about technologies, and lack of (appropriate) technologies were major adoption barriers of SAPs in Kentucky. These results suggest an importance of extension and education to increase adoption of SAPs as well as bring awareness about technologies and practices. The "Happy" attitude of farmers, though not well understood, is another barrier of adoption which requires further research in future.

**Adoption of Sustainable Agricultural Practices among Kentucky Farmers and Their Perception about Farm Sustainability**

**Style Manual and Journal Format:**

Citation and Referencing Style: American Psychological Association (APA) 6th Edition.

Computer Software Used: Microsoft Office 2013, Microsoft Excel 2013, SAS, SPSS.

Master of Science in Environmental Studies, Kentucky State University, May 2017

B. Sc. (Agriculture), Tribhuvan University, July 2013

139 Typed Pages

Major Adviser: Dr. Buddhi Gyawali

**CHAPTER I**

**INTRODUCTION**

The increasing demand of food is causing for the heavy use of chemicals, hybrid and transgenic crops leading to modernizing and globalizing agriculture (Altieri, 2009; Rusch, Valantin-Morison, Sarthou, & Roger-Estrade, 2010). It is creating problems such as environmental degradation, biodiversity losses, disease and pest outbreaks, soil degradation, and ground water level depletion (Altieri, 2009; Feenstra, 2016; Rusch *et al.*, 2010). It has also brought socioeconomic adversities by adding risk on human health and well-being, reduced efficiency and labor demand and thus reducing the viability of agriculture (Altieri, 2009; Bollinger, Hoyt, & Blackwell, 2015; Feenstra, 2016).

The agriculture has always been an important component of Kentucky economy since civil war of 1861 and abolition of slavery. Kentucky ranked as first tobacco producing state until 1929, produced virtually all hemp throughout USA during 1849-1870, Bourbon whiskey, Keenland and Derby have become identity of Kentucky (Gale Encyclopedia of U.S. Economy History, 1999). The Committee of Kentucky (of distinguished citizens) in 1949 ranked agriculture as the
top priority in a report "*Kentucky on the March*" which also
highlights the importance of agriculture since history. In
1940s, 70% of the economy of Kentucky was agricultural.
Agriculture and other agriculture-related industries[2]
contributed about $46.3 billion which is 2% of the state
GDP in 2013. Family farming (90%)being a characteristics of
Kentucky agriculture, is also an integral part of the
society (Bollinger *et al.*, 2015).

Kentucky agriculture is affected by climatic problems
such as severe drought, high temperature, untimely freezes
as well as socio-economic problems such as reduce in farm
incentives, limited labor supply, financial crisis, reduced
market, reduced young farmers (Kentucky Agricultural
Council Task Force on the Future of Agriculture, 2012).
Tobacco Buyout Program (TBP)[3] has brought dramatic

---

[2] All agricultural activities and farms that generates less than
$1,000 revenue per year are considered as agriculture-related industries
(Natural Resource Conservation Service, 2017)

[3] Tobacco Transition Payment Program, commonly known as tobacco
buyout program started in 2005, ended the depression era federal
tobacco price and quota support program since 1938 and started paying
farmers to stop growing tobacco, tobacco quota holders and producers'
transitions to the free market after Fair and Equitable Tobacco Reform
Act, 2004. The buyout program aims to eliminate small, less commercial
and less competitive farmers encouraging farmers who wish to continue structural changes in the agriculture of Kentucky (Kentucky Agricultural Council Task Force on the Future of Agriculture, 2012). Number of tobacco farmers decreased from 56,373 to 8,113 from 1992 to 2007. The Buyout program forced small and marginal farmers to move out of agriculture business (Hart, 2011).

**Table 1**: Characteristics Kentucky Agriculture from 1982 to 2012.

| Year | Number of Farms | Land in Farms (Acres) | Average Farm Size (Acres) | Average Age (Years) | Primary Job as Farming |
|------|------|------|------|------|------|
| **2012** | 77,064 | 13,049,347 | 169 | 57.60 | 32,137 |
| **2007** | 85,260 | 13,993,121 | 164 | 56.50 | 33,935 |
| **2002** | 86,541 | 13,843,706 | 160 | 55.20 | 46,939 |
| **1997** | 91,198 | 13,940,180 | 153 | 53.70 | 36,050 |
| **1992** | 90,281 | 13,665,798 | 151 | 53.20 | 40,175 |
| **1987** | 92,453 | 14,012,700 | 152 | 52.20 | 41,451 |
| **1982** | 101,642 | 14,179,284 | 140 | 50.50 | 49,062 |

(Source: USDA Census Desktop Query Tool. 2012 Census of Agriculture USDA/NASS) (USDA, 2015)

---

growing the crop to increase their crop acreages and production. It also encouraged big farmers to grow as much tobacco they can (Hart, 2011).

Table 1 summarizes statistics of characterisitcs of Kentucky Agriculture as presented by Ag Census 2012. The data supports that the number of farmland and the agricultural land is shrinking down with the increase in average size of farms. The average age of farmers is also increasing every year and has reached the point which was never seen since last 30 years in history. Also the number of farmers who have reported their primary occupation as farming has decreased by 34.50% inbetween last 30 years. The increase in the average age suggests that same generation is continuing farming since long period of time and very few new and young farmers have entered in farmimg. The decreasing trend of operating and relying on farming as a primary occupation suggests decreasing dependency of people on farming and seeking of alternative non-farming jobs. According to Bollinger *et al.* (2015), the decrease in farm numbers is the highest in the United States of America from 2007 to 2012, about half of the farmers works off-farm for more than 200 days/year and the share of agriculture in Kentucky economy has declined over several years.

Farmers staying in farming business are diversifying their farms incorporating horticulture enterprise (Bollinger *et al.*, 2015; Smith, Altman, & Strunk, 2000), agrotourism (Brown & Swanson, 2014), high value crops (Hull, 2002), livestocks, fishes, timbers, ranches (Bollinger *et al.*, 2015). Agro-tourism (Brown & Swanson, 2014), alternative high value crops (Hull, 2002) are becoming new farming strategies in Kentucky. The interest in sustainable organic farming practices, agriculture diversification, and alternative marketing such as Community Supported Agriculture (CSA) and Farmers market have also increased (Tanaka, Williams, Jacobsen, & Mullen, 2012).

Farmers, however, are provided with limited alternatives and substitutes in comparison to lucrative tobacco as well as virtually no option for marketing and processing infrastructure to support diversification (Smith *et al.*, 2000). Despite constraints, farmers are identifying new and diversified farming strategies by adding alternative enterprise, i.e. adding values. These strategies are providing new economic opportunities, enriching farm sustainability as well as, supporting the rural economy (Bollinger *et al.*, 2015).

The awareness about crop diversification (Smith *et al.*, 2000), crops and livestock integration, new farming strategies, ecological harmony (Feenstra, 2016; Scherr & McNeely, 2007), health and environmental concerns among farmers (Bollinger *et al.*, 2015; Feenstra, 2016) is increasing. Policy makers are encouraged to develop alternative, attractive and sustainable agricultural system supporting long term profitability, stewardship of natural resources, and maintain quality of life (Feenstra, 2016). These changes discussed above suggest the need assessment reconsidering the relationship between agro-ecosystem and natural ecosystem (Scherr & McNeely, 2007) to meet the current food demand and supply through holistic food system approach (Welch & Graham, 1999). The sustainable agriculture development represents a new approach in the agriculure that ensures the economic and social beneifts for the current generation without comprimising the demand of future generation without disturbing the ecosystem (Zaharia, 2010).

**Statement of Problem:**

Designing alternative sustainable food system, which is locally adaptable, meeting local demands and attractive to farmers, requires assessment of farmers' perception about achieving sustainable objective. Sustainability assessment is the starting key point to develop innovative technologies and sustainable farming system (Lebacq, Baret, & Stilmant, 2013). Understanding of variation of farmers affinity towards adoption of differnet technologies and practices requires understainding of variability of farmers' sociodemographic, geographic, and farm attributes (Kornegay *et al.*, 2010). Adoption of sustainable agricultural technologies, although, is common in Kentucky, systematic study to understand sustainability practices and technologies adoption, their governing factors as well as adoption barriers are less understood. This research aims to fill the gap by exploring farmers' perception about their farm sustainability, variations in adoption of SAPs and their adoption barriers.

**Objectives and Hypothesis**

**Objectives:**

   The specific objectives were:

     1) to explore Farmer's Perceptions about Farm and
       Farming Practice Sustainability

     2) to identify predictors of SAPs adoption using farm
       attributes, farmers' attitudes and behaviors,
       socioeconomic and demographic factors, and
       knowledge.

     3) to evaluate adoption barriers of SAPs among Kentucky
       Farmers.

**Hypothesis:**

   The adoption of SAPs among Kentucky farmers have
significant relationship with their farm attributes,
farmers' attitudes, socioeconomic, and demographic factors.

**CHAPTER II**

**LITERATURE REVIEW**

**Sustainable Agriculture: Definition, Concept and Objectives:**

The word "sustain' is derived from the Latin word *sustinere* (sus-from below and tenere-to hold) to keep in existence or maintain, implies long term support or permanence (Zaharia, 2010). The Long term sustainability of agriculture and food system has been a concern for scientists since long time which can be traced back to the environmental degradation during 1950s-1960s (Pretty, 2008), though, the concept became prominent after Brundtlands' Report in 1987 (Velten, Leventon, Jager, & Newig, 2015).

The US congress (1990) defined sustainable agriculture as a form of 'practices' which is commonly being adopted by stakeholders involved in sustainable agriculture sector (Dakers, 1992, Liaghati, Veisi, Hematyar, & Ahmadzadeh, 2007) including United States Department of Agriculture (USDA) (Gold, 2007). US congress (1990) (Farm Bill, Public Law 101-624, Title XVI, Subtitle A, Section 1603, P: 1240) defines: "the term sustainable agriculture means an integrated system of plants and animals production practices having a site-specific application that will, over the long-term a) satisfy human food and fiber needs; b) enhance environmental quality and the natural resources base upon which the agricultural economy depends; c) makes the most efficient use of non-renewable resources and on-farm resources and integrate, where appropriate, natural biological cycles and controls; d) sustain the economic viability of farm operations; and e) enhance the quality of life for farmers and society as a whole."

Sustainable agriculture is a multi-dimensional concept and a holistic approach of sustainable development with the goals of achieving environmental health, economic profitability and social and economic equity (Douglass, 1984; Kornegay *et al.*, 2010; Sands & Podmore, 2000). The agriculture sustainability should extend both temporally and spatially, globally valuing the welfare of current and future generation of all species living in the biosphere. It should address the issues of entire food system nexus in local, regional, national, and international level (Allen, Dusen, Lundy, & Gliessman, 2013).

Sustainable agriculture describes farming systems that are capable of maintaining their productivity and usefulness to society in long run which also must be resource-conserving, socially supportive, commercially competitive, and environmentally sound (Gold, 2007; Liaghati *et al.*, 2007). The agriculture is directly influenced by the interaction among environemental, social, and economic aspects of a location and thus should maintain the balance between these aspects (Allen *et al.*, 2013). The sustainable agriculture should be environmentally sound, economically viable and feasible, socially just and responsible, culturally acceptable, non-exploitative and serve as a foundation for the future generation (Allen *et al.*, 2013; Filho, Young, & Burton, 1999; Reijntjes, Haverkort, & Waters-Bayer, 1992). It should seek to optimize management, reducing external inputs as well as farm resources without harming environment, public health, human communities as well as animal welfare (Reijntjes *et al.*, 1992).

Sustainable agricultural (or agriculture) practices and sustainable agriculture are often used together. The sustainable agriculture is a long term concept with the goal of achieving environment health, economic profitability, socioeconomic integrity and equity. Agriculture sustainability should integrate three goals of environmental health, economical profitability, and social and economic equity (Douglass, 1984; Kornegay *et al.*, 2010; Sands & Podmore, 2000). Sustainable agriculture should focus on the development of technologies/practices that support the goal of enhancing environmental quality, goods, services and agriculture productivity. It also should maintain the economy viability, fulfill the basic needs of food, fiber, feed, environmental goods, services and agriculture productivity (Kornegay *et al.*, 2010; Pretty, 2008).

Sustainability in agriculture can be achieved through the adoption of sustainable agriculture practices and farming techniques. The sustainability in agriculture can be achieved by maintaining the ecosystem capacity, efficient use of natural resources, human capacities and knowledge, energy resources, equitable distribution of goods supplied as well as the negative consequences of agricultural development (Zaharia, 2010). Sustainable agriculture adopts productive, competitive and efficient practices, while protecting and improving the environment and the global ecosystem, as well as the socio-economic conditions of local communities (Hani, 2006). These practices, together, should form a farming system that meets the agricultural demand as well as conserve natural resources and maximize net benefit of ecological services and functions (Kornegay *et al.*, 2010).

**Sustainable Agricultural Practices:**

Sustainable agricultural practices have identified by various names such as Low External Input Sustainable Agriculture (LEISA) (Filho et al. 1999; Reijntjes *et al.*, 1992), and Best Management Practices (BMPs) (Baumgart-Getz, Prokopy, & Floress, 2012). The fundamental characteristics of the concepts include use of local resources optimally (Coze & Hedrich, 2007; Mesner & Paige, 2011), economically feasible, ecologically sound, culturally adopted, and socially just (Filho *et al.*, 1999; Reijntjes *et al.*, 1992), reduce risk from crop failure, reduce cost of production increasing long and short term benefits (Najafabadi, Khedri, & Lashgarara, 2012). All of these concepts have common objectives of achieving least possible adverse impact in the environment, human, animal or plant heath by utilizing local resources optimally to improve livelihood of farmers and the community (Kornegay *et al.*, 2010; Mesner & Paige, 2011).  They are methods or practices which, when installed or used, are consistent with efficient, practical, technically and environmentally sound animal or crop production practices and are best suited to preventing reducing or correcting agriculture-related problems (Coxe & Hedrich, 2007).

Sustainable agricultural practices are not, necessarily, the set of properly defined practices but are more localized and site specific (Coxe & Hedrich, 2007; Mesner & Paige, 2011; Rankin, 2015). The practices should fulfill the site specific need in specific situation rather than solving a broad problem (Rankin, 2015). For example, small mammals, which loosen the soil, are found to be beneficial in reestablishing and enhancing the vegetation succession in coal mining areas of eastern Kentucky (Larkin *et al.*, 2008). However, it should not be confused that practices suitable in one region may equally adoptable based on its usefulness and availability. The practice should not only be locally available but also use local resources efficiently (Lashgara, 2011). They should provide appropriate technology to solve farmers' problems. It should empower farmers to make choices about how the practices should be used and be accompanied with the available information to make decisions (Rankin, 2015). Nonetheless, the practices should be integrative and holistic addressing issues of agriculture, environment, ecology, life quality, society, economy and food nexus. The interrelationship among SAPs and its characteristics are visualized in the figure below (Fig.1).

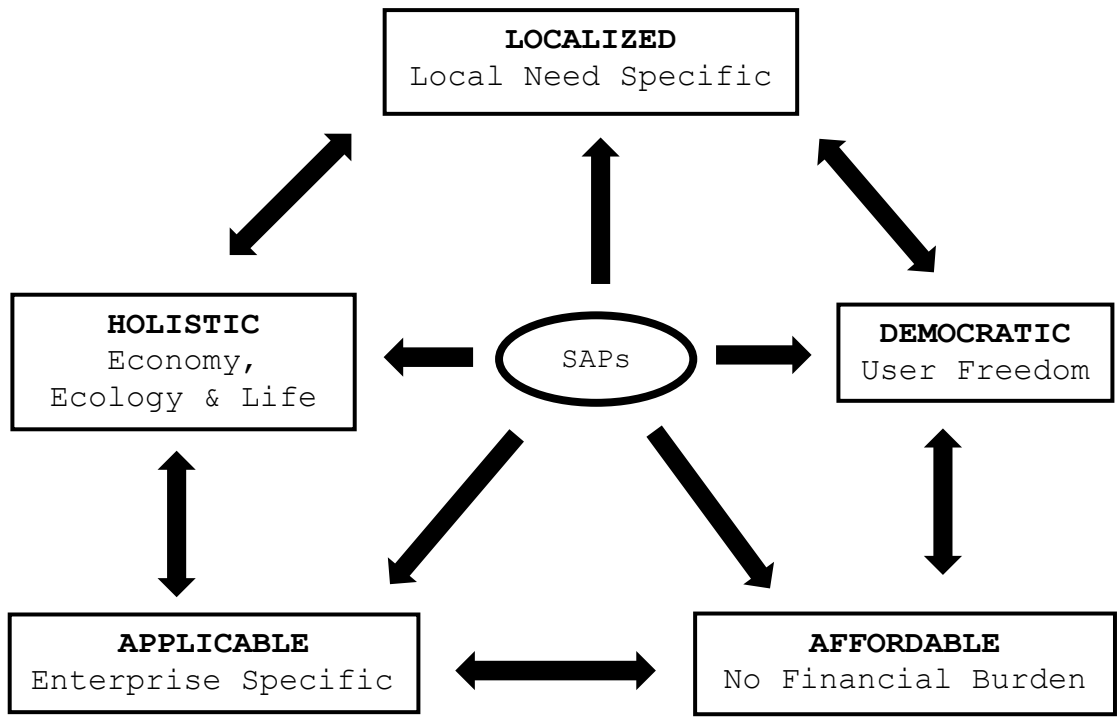

**Figure 1:** Interrelation between Sustainable Agriculture Practices and Their Characteristics.

The SAPs seek to optimize management, reducing external inputs, as well as use local farm resources, however, it does not exclude external inputs but incorporate as complementary to the local resources meeting five criteria above (Fig. 1.) (Reijntjes *et al.*, 1992). They are designed to produce least possible adverse impacts on the environment, human, animal or plant heath (Coxe & Hedrich, 2007; Mesner & Paige, 2011) as well as protect the biological sources (Lashgarara, 2011).

**Factors Affecting Adoption of Sustainable Agricultural Practices:**

The "adoption" process includes five stages: awareness, interest, assessment, examination and adoption. Adoption, broadly, is affected by two factors: individual and structural. The individual aspects are factors such as knowledge, experience, economic state (such as farm size, income), personal characteristics (such as age, sex), needs, motivations, and reference group. The structural factors are socioeconomic conditions environmental requirements (such as population) and communicative systems (Lashgarara, 2011).

Farmers decision to use certain farming practices and their attraction towards sustainability are influenced by many factors such as knowledge, skills, market, public policies, and their own values, resources and land tenures. Market, policy and institutional context are important drivers in US agriculture but response of farmers could be diverse (Kornegay *et al.*, 2010). Among Michigan farmers, the adoption of new environmental-friendly management practices depended upon awareness, attitudes, available resources, and incentives. Farmers are unwilling to adopt low-input practices due to their perception of lower profitability and higher labor requirements (Swinton, Rector, Robertson, Jolejole-Foreman, & Lupi, 2015). The probability of farmers adopting SAPs decreased by the increase in farm size and increase with the involvement in the farmers' organization, contact with non-governmental organization, awareness with negative effect of chemicals on health and environment, reliance on family labor, and farming with better soil condition (Filho *et al.*, 1999). A research among Eastern Uganda small farmers, however, shows that the average size of farm has positive relation with the use of erosion control agricultural practices (Barungi *et al.,* 2013). The diffusion of a sustainable practice was correlated with the reduction in profitability of agriculture, increase in price of farm inputs, and relatively cheaper farm labor among Brazilian farmers (Filho *et al.*, 1999).

The adoption of conservation practices is also affected by tenure, age, knowledge and attitude towards a program (Kabii & Horwitz, 2006). It is also found that renters are less likely to adopt soil health related agriculture management practices compared to the owner of the land (Bergtold, Duffy, Hite, & Raper, 2012; Carlisle, 2016; Upadhyay, Young, Wang, & Wandschneider, 2003). However, share renters'[4] behavior was more inclined towards adoption of conservation compared to cash renters (Carolan, 2005; Soule, 2001).

The adoption of conservation tillage is correlated with the (college) education (Fuglie, 1999; Soule, 2001; Upadhyay *et al.*, 2003 & Wu and Babcock, 1998). In contradiction, a meta-analysis of articles published in US from 1980 to 2005 found that socio-economic, demographic attitude and awareness related variables were non-significant in majority of the studies (Prokopy, Floress, Klotthor-Weinkauf, & Baumgart-Getz, 2008).

Younger age—as older age has shorter horizon of planning than younger farmers— (Altman, et al., 1998; Baumgart-Getz *et al.*, 2012) and college education were positively related with the adoption of on-farm activities to supplement the tobacco farming. These actions and perceived barriers of adoption are also largely dependent upon college education, off-farm income, and large farm size, percent income from tobacco, and age (Altman *et al.*, 1998).

---

[4] Renting process includes sharing of farm product as well as distribution of risk of farming among renters and landlords (Carolan, 2005).

Baumgart-Getz *et al.* (2012) identified capital as the best financial predictor of adoption behavior. Carlisle (2016) further mentioned financial barriers as an important decision making factor for the adoption of soil health related practices. However, immediate financial benefits are less important compared to long-term soil health, if practices have been already adopted. US farmers adopted improved seed more rapidly compared to conservation tillage practice (Lambert, Sullivan, Claassen, Foreman, 2006) as conservation tillage require different set of tools which is wise to adopt only when old technologies are ready to be replaced (Krause and Black, 1995).

Baumgart-Getz *et al.* (2012) found that access to quality information, financial capacity, connection to local agency and farmers' network and watershed groups have large and significant effect in the adoption of best management practices in the United States. Baumgart-Getz *et al.* (2012), however, assert that formal education has non-significant and extension training has positive influence in the adoption of agricultural practices.

Coughenour (2003) found that the spread of no-tillage in Kentucky is due to the interaction and networking among farmers, cropland and agriculture advisors, company representatives and agriculture scientists. The networking among farmers and agencies have large influence in farmer's decision to adopt the practices (Carlisle, 2016; Coughenour, 2003). However, networking could play a role in both, negative and positive directions: networking with organic fertilizer dealers negatively influence the adoption of cover crops in organic farms in several states in the US (Gabrielyan, Chintawar, & Westra, 2010). Same variables, however, could not explain the adoption process consistently throughout the world (Knowler and Bradshaw, 2007).

Carlisle (2016) found that the adoption of soil health related management practices (cover crops, crop rotation and conservation tillage) are affected by financial issues (initial investment, opportunity cost of cash crops ongoing investment in factors of productions), agronomic issues (cropping seasons, establishment and germination), land tenures, government policies such as Farm Bill Conservation Program (Baradi, 2009; Coughenour 2003; Fuglie, 1999; Soule; 2001; Wu and Babcock, 1998).

Farmers' pervious participation in the conservation program has positive influence in the adoption of additional long-term conservation practices (Featherstone and Goodwin, 1993; Bergtold *et al.*, 2012; Lichtenberg, 2004; Singer, Nusser, & Alf, 2007; Upadhyay *et al.*, 2003; Wilson, Howard, & Burnett, 2014). The chains of practices created following the previous one (also called a "foot in the door" model) could lead to the complete transformation of farming systems (Wilson *et al.*, 2014). Farmers, however, in Missouri are discouraged to transport manure due to its odor, undigested weed seeds and economic issues (Battle & Krueger, 2005).

**Adoption Barriers of Sustainable Agricultural Practices:**

Swinton *et al.* (2015) and Nowark (1992) found that farmers don't adopt the sustainable agricultural practices[5] due to two major reasons or both: unwillingness and unaffordability. The technologies, which are cost effective and have wider applicability would spread more easily and widely compared to technologies that are cost ineffective and less applicable. Cost effective and advantageous technologies are rapidly adopted by farmers and costly technologies, though they are attractive, are relatively adopted slowly. However, large farmers, in contrast, adopt capital-intensive and attractive technologies more rapidly in comparison to small and resource poor farmers (Lambert *et al.*, 2006).

The success of the practices is dependent on the selection, design, and implementation of technologies as well as, appropriate evaluation of farm resources, operation, farmers existing practices and needs and market structure (Coxe & Hedrich, 2007; Mesner & Paige, 2011).
Farmers tend to apply the practices and technologies, which
can be quickly and easily tested on their farm land
(Carlisle, 2016).

---

[5] Practices mentioned by Swington *et al.* (2015) are small grains in crops rotation, conservation tillage, fertilizer reduced rate, soil nitrogen testing on corn, cover crops, legume crops, scout for pest, reduced tillage, wheat in corn-soy rotation, manure application, no tillage (for some years and more than 4 years), reduced rate of N, herbicides and insecticides.

Among Iowa farmers, females tend to have more positive
attitude towards conservation compared to men but have less
knowledge and thus less likely to adopt conservation
practices (Druschke & Secchi, 2014). The adoption of a
practice or technology, however, does not necessarily means
using it in the farm. It was found that though the adoption
of soil erosion control related practices is high among
Uganda farmers, their use is moderate. Besides, a large
number of farmers are using technologies in small scale
(Barungi *et al.*, 2013).

Few researches identified slope and erodible land have
positive correlation with the adoption of conservation
practices (Fuglie, 1999; Soule, 2001; Wu and Babcock, 1998).
Farmers are interested in diversifying their farm operation
by adding high value and short season crops (Lambert *et al.*,
2006; Lichtenberg, 2004; Singer *et al.*, 2007; Snapp *et al.*,
2005).

Identifying the role of various socioeconomic, biological, demographic, farm and other attributes on the adoption decision has become more important recently due to changing demographics and regulations. As said by Knowler and Bradshaw (2007), even a single variable cannot explain the phenomenon universally. The effect of same variable may differ based on the location as well the types of society and farmers. This calls for further research on understating the site specific effects of variables and their impacts on farmers' decision making process.

**CHAPTER III**

**MATERIALS AND METHODS**

This chapter describes the study site, and methods for sampling, data collection, and analysis.

**Overview of Study Area:**

Kentucky has diverse agriculture, as its landscape, from the bottom of Mississippi river in the east to the Appalachian Mountains. Majority of counties are agricultural with each county having at least some forms of agriculture (Bollinger *et al.*, 2015). The state comprises of 117 rural counties out of 120, with its rakings in bottom 6th position in poverty. Kentucky holds large number of small and family farmers with horses, beef cattle, poultry, row crops, hay crops, vegetables, horticulture and nursery crops as major crops (Hancock, 2015). This study covers entire state of Kentucky.

**Sampling Design, Survey Dispatch and Data Collection:**

Survey questionnaires were pretested among farmers in Third Thursday Thing[6] at Kentucky State University as well as with other individual farmers by USDA/NASS in Louisville. Feedbacks from farmers were addressed before finalizing the survey. A final copy of the survey (only relevant) is attached in Appendix F.

Sampling of farmers for survey were done using the database of farmers maintained by National Agriculture Statistics Service of United States Department of Agriculture (NASS/USDA). Double stratified sampling method was used to sample surveying farmers. The total value of sales of a farm was divided into four groups and cross stratified by six agriculture districts of Kentucky making 24 strata throughout the state. The programming was done to put 25% of the cumulative total measure of size of variable into four strata, which resulted into 15%, 31% 39% and 15% of total survey allocated to strata 1, 2, 3 and 4 respectively (J. Stephens-NASS/USDA, Mail Interaction,

---

[6] "Third Thursday Thing" is an agriculture extension program which is held every third Thursday of each month at Kentucky State University Research and Demonstration Farm with Kentucky farmers.

August 31$^{st}$, 2016). Easiness of data collection and budget constraints were taken into consideration during stratification process. Categories of each stratifying variables were provided in Appendix B.

Surveys were mailed to 1,000 farmers from the North Carolina Print Mail Center and returned and documented at Regional and Field Office (RFO) of USDA/NASS, Louisville, Kentucky. The number of samples dispatched in each strata is provided in Appendix A. Farmers were requested to report information of their farms and farming details for the year 2014. Farmers were followed up using phone calls which was helpful to improve quality and quantity of the survey. The majority of the surveys were conducted in-between September 10, 2015 and November 27, 2015. Large number of phone calls were made during October, 2015 and data collection was closed on January 13, 2016 (J. Stephens-NASS/USDA, Mail Interaction, August 31$^{st}$, 2016). Out of total 1,000 survey dispatched thoroughout Kentucky, 207  responded through the mailed survey whereas 458 farmers were interviewed using phone calls and 335  farmers were either inaccessible, non-responders or refusers. Out of total 230 respondents, 97

(42.20%) of farmers responded the survey by phone whereas
133 (57.80%) of farmers responded the mailed survey.

**Data Processing, Quality Monitoring and Preparation for Analysis:**

Mailed and Telephone survey were recorded in a hard copy format and cases were numbered from 1001 to 2000 to organize the data. The survey was transformed into Microsoft Office 2013 using the numbering of the cases in the copy and were extracted to Microsoft Excel using SAS[7] in NASS/USDA data lab in Louisville.

Surveys were randomly cross verified by USDA/NASS staff by cross checking information, and also tallied individual cases to detect any unusual response. The demographics and farm attributes were further cross-checked with respective information of Kentucky from US Ag Census 2012 And found that the statistical parameters of survey data (demographic attributes) are close to that of the Ag Census data.

---

[7] SAS is a statistical software developed by SAS Institute for Advanced Analytics, Multivariate Analysis, Business Intelligence, Data Management, and Predictive Analysis.

Data analysis was conducted in IBM SPSS[8] 24.0 using 230 cases out of 1,000 surveys. The variable "Land Operated" had 100% response therefore was used as a benchmark to select 230 cases for further analysis. Farmers who reported the "Land Operated" variable have also reported majority of the questions in the survey.

**Post-Strata Weighting:**

Post stratification is a common technique in survey for incorporating population distribution of variables into survey estimates. The basic technique divides samples into post-strata and computes post-stratification weight $W_{ih} = rP_h/r_h$ for each sample case in the post-stratum $h$, where $r_h$ is the number of survey respondents in the post stratum $h$, $P_h$ is the population from a census, and $r$ is the respondent sample size (Little, 2012). The proportional weight is used for this study which is provided in a simplified form below (Eq. 1).

---

[8] IBM SPSS 24.0 is the latest version of Statistical Package for the Social Science (SPSS) software developed by International Business Machines Corporation (IBM) used frequently by social science researchers and diverse institutions in data analysis.

Proportional Weight:

$$Weight = \frac{Proporiton\ of\ Population}{Proportion\ of\ Sample} = \frac{\dfrac{Population\ (N)\ of\ a\ strata}{Total\ Population\ (\sum N)}}{\dfrac{Sample\ (n)\ in\ the\ strata}{Total\ Sample\ (\sum n)}}$$

Eq. 1 (Maletta, 2007).

**Data Analysis:**

This thesis has three objectives which are discussed below separately:

**Objective I: Farmer's Perceptions about Farm and Farming Practice Sustainability:**

Farmers' perception of farm sustainability was explored using the rating scale data. Farmers were requested to rate their perception about their farm meeting or implementing the farm level sustainable agriculture objectives viz. Local Ecosystem Sustainability, Social Sustainability, Social Acceptability, and Food System Sustainability in five rating scale—1 being "Not Sustainable" to 5 being 'Very Sustainable'. General frequency count was used and interpreted for this objective.

**Objective II: Identify Predictors of SAPs Adoption using Farm Attributes, Farmers' Attitudes and Behaviors, Socioeconomic, and Demographic Factors and Knowledge:**

Focus group and previous study-based sustainable practices were listed in survey questionnaire giving option for farmers to choose whether they have adopted sustainable agricultural practices—planning to adopt or not applicable—or not adopted in their farm in designated year. Total number of sustainable practices adopted by respondents was considered as the dependent variable in the regression which is a count variable. The dependent variable was computed by counting number of sustainable agricultural practices marked as "adopted", among listed practices, by each of the respondents.

**Model Specification: Poisson Distribution and Poisson Regression Model:**

Poisson regression is a Generalized Liner Model (GLiM), based on the Poisson distribution representing the distribution of error, which provides accurate results for the data set having non-negative count integer dependent variables. The Poisson distribution is a member of a set of *exponential family* probability distribution: the height of the probability curve for specific value of *y*, called the *probability density*, contains an exponential function (Coxe, West, & Aiken, 2009). The probability density (or the height of the curve) depends upon two parameters, mean (μ) and the variance (σ$^2$) in normal distribution curve. However, the Poisson distribution differs from normal distribution therefore, Poisson regression is more attractive to represent count data (Coxe *et al.*, 2009).

The equation for normal distribution is

$$f\left(\frac{y}{\mu}, \sigma^2\right) = \frac{1}{\sqrt{2\pi\sigma}} e^{-\frac{(y-\mu)^2}{2\sigma^2}}$$    Eq. 2.

(Coxe *et al.*, 2009, p.122).

The probability mass function for the Poisson distribution (Eq. 3) gives the probability of observing given value, *y*, of the variable Y that is distributed as a Poisson distribution with the parameter μ—the athematic mean of number of incidents that occurs in specific time interval. The Poisson distribution would yield the probability of 0, 1, 2, 3 … incidents, given the mean (μ) of the distribution.

$$P\left(Y = \frac{y}{\mu}\right) = \frac{\mu^y}{y!}e^{-\mu} \qquad\qquad\qquad \text{Eq. 3.}$$

where, y! = y-factorial = y x (y-n)! = y (y-1) (y-2) … (2)(1).

(Coxe *et al.*, 2009, p.123).

The probability of a specific count also depends on the variance ($\sigma^2$) of the number of counts. However, the poison distribution has only one parameter $\mu$ i.e. both the conditional mean and the conditional variance are equal to $\mu$, the condition is called as *equi-dispersion* (Coxe *et al.*, 2009; Avei, Alturk, & Soylu, 2015). This property of Poisson distribution, i.e. the variance increases with the increase in mean and vice versa, is helpful in modeling count outcome (Coxe *et al.*, 2009).

**Model for Over-dispersed Data and Negative Binomial Regression:**

The simplest adjustment for over-dispersion in the Poisson regression includes second parameter used in the estimation of the conditional variance known as *over-dispersion scaling parameter, ($\phi$)*. The corrected model assumes Poisson error distribution with mean $\mu$ and variance $\phi\mu$. The scaling parameter *($\phi$)* will be greater than 1, if over-dispersion is present in the model, 1 in equi-dispersion and below 1 if under-dispersion making the resulting model equivalent to Passion distribution (Coxe *et al.*, 2009). This is also called as Quasi-Poisson methods (Avei *et al.*, 2015). The dispersion in the model is determined by the Pearson chi-square ($\chi^2$) goodness-of-fit statistics (Eq. 4):

$$\phi = \frac{\chi^2 \text{Pearson}}{\text{df}}.$$

Eq. 4.

(Coxe *et al.*, 2009, p.131).

The over-dispersed model allows conditional variance to be larger than their corresponding means making standard error larger than standard error of Poisson model by a factor of $\sqrt{\phi}$ and the interpretation of the coefficient is ideal to standard Poisson model. The deviance for the model is adjusted by the scaling factor and the deviance for the over-dispersed Poisson model is equal to the deviance of standard Poisson model divided by $\phi$. The smaller deviance of the model indicates a better fit (Coxe *et al.*, 2009).

Negative Binomial model is a popular choice and suggested as an alternative of the Poisson regression for the over-dispersed data (Berk & MacDonald, 2008; Coxe *et* *al.*, 2009; Aevi *et al.*, 2015). The negative binomial model accounts over-dispersion by assuming that there will be unexplained variability among individuals who have the same predicted value. This additional unexplained variability between individuals leads to larger variance (than expected by Poisson distribution) in overall distribution but has no effect in mean. This additional variability is conceptually similar to the inclusion of an error term in normal linear regression (Coxe *et al.*, 2009).

The Negative Binomial distribution equation for $(y_i|x_i)$ arise as a Gamma mixture of Poisson distribution. One parameterization of its probability density function is

$$f(y; \mu, \theta) = \frac{\Gamma(y+\theta)}{\Gamma\theta y!} \frac{\mu^{y}\theta^{\theta}}{(\mu+\theta)^{y+\theta}} \qquad \text{Eq. 5.}$$

with mean $\mu$ and shape parameter $\theta$; $\Gamma(.)$ is the Gamma function. For every fixed $\theta$, this is another special case of the GLiM framework. It also has $\phi = 1$ but with variance function $V(\mu) = \mu + \frac{\mu^2}{\theta}$. The mean of negative binomial regression can also be assumed to follow the log link $E(Y_i = \mu_i = \exp(x_i'\beta),$ and the maximum likelihood estimates can be obtained by maximizing the log likelihood (Avei *et al.*, 2015, p.2).

Poisson regression is a GLiM with Poisson distribution error structure and the natural log (ln) link function. The predicted score in the Poisson regression is not itself a count, but the natural logarithm of the count. Hence, Poisson regression is "Linear in the Logarithm". Indeed, all form of GLiM has same property—there is a form of the regression equation that is linear (Coxe *et al.*, 2009).

The Poisson regression model can be depicted as

$$\ln(\hat{\mu}) = b_o + b_1 X_1 + b_2 X_2 + \ldots + b_n X_n + \varepsilon \qquad \text{Eq. 6.}$$

where, $\hat{\mu}$ is the predicted count on the outcome variable, $b_o$ is the intercept of the regression model and $b_1$, $b_2$, …, $b_n$ are coefficient of respective predictors $X_1$, $X_2$, …, $X_n$ and $\varepsilon$ is an error term (Coxe *et al.*, 2009, p.123).

**Interpreting the Model and Coefficients:**

One interpretation of the regression coefficient is in terms of $ln(\hat{\mu})$ i.e. a 1-unit increase in the $X_1$ results in a $b_1$ unit increase in the $ln(\hat{\mu})$, provided other variables are constant in the model. But, researchers would like to measure the effect of predictors in the number of times the event occurs. The second interpretation can be obtained by algebraic manipulation i.e. raising power $e$ in both side of equation 6, which becomes

$$e^{\ln(\hat{\mu})} = e^{b_o + b_1 X_1 + b_2 X_2 + \cdots + b_n X_n} \qquad \text{Eq. 7.}$$

A property of $e$ and the natural log is that the $e^{\ln(x)} = x$, which gives

$$\hat{\mu} = e^{(b_o + b_1 X_1 + b_2 X_2 + \cdots + b_n X_n)} \qquad \text{Eq. 8.}$$

A property of exponents is that $x^{a+b+c} = x^a x^b x^c$ which separates above equation into small parts, resulting into an equivalent equation:

$$\hat{\mu} = e^{b_o} e^{b_1 x_1} e^{b_2 x_2} \dots e^{b_n x_n} \qquad \text{Eq. 9.}$$

Or,

$$e^{b_1 (X_1 + 1)} = e^{b_1 X_1 + b_1} = e^{b_1 X_1} e^{b_1} \qquad \text{Eq. 10.}$$

The $e^{b_1}$ term in the equation 9 is the effect of a 1-unit change in $X_1$ on the outcome. For 1-unit increase in $X_1$, the predicted count ($\hat{\mu}$) is multiplied by $e^{b_1}$, holding all other variables constant in the model (Coxe *et al.*, 2009, p.124).

**Objective III: Evaluate Adoption Barriers of SAPs among Kentucky Farmers:**

Barriers to adoption of sustainable agricultural practices are discussed in the third objective. Descriptive statistics were used to analyze the barriers of the adoption.

**CHAPTER IV**

**RESULTS AND DISCUSSION**

The result section provides a discussion of various socioeconomic, demographic characteristics of respondents, farm attributes, as well as three objective specific results.

**Farmers' Characteristics and Farm Attributes:**

Farmers with less than $350,000 Gross Cash Farm Income (GCFI) are considered as small farms (Hoppe & MacDonald, 2016). Among 230 respondents, 77.90% of farmers have annual income less than $250,000 with 41.30% of farmers below $10,000 per year. Only 3.30% of famers have annual income above $250,000 (Appendix C, Table 10). More than three-fourth farmers with income below $250,000 shows that the survey results are more inclined towards predicting the behavior of small farmers. Also, the survey was dominated by White farmers which was followed by African American farmers. Other races identified were Native American or

Alaska Native and two or more races, however their
responses were negligible.

All businesses that have "sole proprietorship" or
proprietorship under "family held corporation" are
classified as family farm[9] in this report. 74.20% farmers
reported they individually owned the business whereas 3.10%
have proprietorship under family held corporation making
77.80% of farmers under family farming throughout the state
of Kentucky. Other types of ownerships reported in the
survey were partnership (5.8%), trust or estate (3.0%)
(Appendix C, Table 9).

The average age of respondents was 62.85 ± 12.24 years.
Though the average age is higher than that reported by Ag
Census 2012 (57.60 years) (National Statistics, 2015), the
survey data follows the trend. Ag Census 2012 data shows
that the average age of Kentucky farmers is highest in the
last 30 years which was never observed before. More than two third respondents (71.80%) were at least 50 years, more than half (53.90%) are above 60 years and none of the respondents were below 30 years old. Only 18.20% of the respondents belongs to the age group between 31 and 50 years and 11% respondents were beginning farmers[10]. The respondents were involved in farming business and taking major farming decision since 33.28 ± 18.49 and 29.86 ± 16.34 years in an average respectively. Majority of farmers were involved in farming for at least 20 years and making farming decision since at least 21 years to maximum 60 years (Appendix C, Table 9).

---

[9] Hoppe and McDonald (2016) defined any farm where the operator
and persons related to the operator do not own a majority of the
business as nonfamily farm. The thesis considers family farm if the
majority of the business is owned by operator and person related to
operator.

The formal education level of majority of Kentucky farmers is between high school and college degree. 13.0% respondents have "less than high school" degree, 30.80% have high school degree, and 19.10% have some college level education. 11.20% and 9.00% of the respondents have the college degree and graduate degree, respectively. Also, more than one third (37.40%) of farmers' spouse have below college degree of formal education (Appendix C, Table 11).

---

[10] Farmers or entity who has not operated a farm or ranch or who has operated farm for not more than 10 consecutive years (Natural Resource Conservation Service, 2017).

The average farm land operated, owned and rented are 169.60, 144.60 and 197.58 acres respectively. Half of the farmers reported less than 100 acres of operated land and more than 80% respondents reported below 250 acres. Similarly, 53.50% of respondents owned less than 100 acres of farm land and 83.60% farmers own less than 200 acres of land (Appendix C, Table 11). The average of total land rented is much higher than total land operated as well as total land owned. Also USDA Farm Service Agency (FSA) Transition Incentive Program (TIP) is encouraging beginner or socially disadvantaged farmers to rent the farmland instead of buying it (Farmland Information Center, 2017).

Among the total 230 respondents, 65.80% farmers have grown row crops in their farms, 16% reported they have vegetables and 77.50% have livestock. The sum of percentage of all three types of farmers is more than 100%, which signifies some but not all farmers have integrated at least one of these three enterprises in their farms. This is also supported by the result that 37.10% of farmers, among 230, are in favor for diversifying their farm.

**Objective I: Farmer's Perceptions about Farm and Farming Practice Sustainability:**

Social acceptability of the agricultural practices is largely determined by social norms, values, attitudes and behavior of members of the society (Rankin, 2015). Farmers were provided questionnaire with rating scale for farm level objectives—Local Ecosystem Sustainability, Social Sustainability, Social Acceptability and Food System Sustainability—of sustainable agriculture and rating scale from 1, being Not Sustainable to 5, being very sustainable for each of the objectives. Farmers were asked to rate the sustainability of their farm and farming practices from 1 to 5 for each of the above mentioned objectives.

Local ecosystem sustainability in this research deals with the soil, water, air, biodiversity, and critical species of the locality and impact of agriculture in the components of ecosystem. Food system sustainability refers to the food quality, nutrition, affordability and
accessibility of food by the consumers as well as
producers. Social acceptability refers to the connection of
practices with the rules, regulation, norms social values
and moralities of the communities. Social sustainability is
concerned with human well beings, quality of life of
farmers and labor, labor health issues, economics (Kornegay
*et al.*, 2010).

**Table 2:** Rating of Farm and Farming Practices Sustainability in Kentucky.

| Sustainable Agriculture Objectives | Objectives Rating Scale (1= Not Sustainable  5 = Very Sustainable) | | | | |
|---|---|---|---|---|---|
| | **1** | **2** | **3** | **4** | **5** |
| Social Sustainability (N = 131) | 13 9.50% | 13 10.00% | 34 25.60% | 49 37.10% | 23 17.80% |
| Social Acceptability (N = 134) | 11 8.40% | 15 11.30% | 24 17.70% | 58 43.50% | 26 19.10% |
| Food System Sustainability (N = 133) | 17 12.50% | 11 8.10% | 27 20.50% | 54 40.70% | 24 18.10% |
| Local Ecosystem Sustainability (N = 144) | 16 11.30% | 4 2.50% | 24 16.50% | 53 36.80% | 47 32.80% |

More than 50% of farmers have rated their farming practice 4 or above for all of the four objectives of sustainable agriculture and roughly, 20% of farmers have rated 2 or below for all sustainable agriculture objectives in a one-five rating scale. The survey was provided with few clue words in each of the objectives to help them to understand about various aspects covered by each objective. Those objectives, however, themselves, cover broad elements of sustainability objectives and therefore, results may be taken as a general understanding of farmers.

Since, majority of the respondents have education level below college degree, we can expect that large proportion of farmers might not be very familiar with the objectives of sustainable agriculture. Moreover, farmers may not have proper understanding of agriculture sustainability with the information provided. Therefore, this result should be taken as more general indicator of the perception about sustainability of their farm and farming activities. More detail breakdown of each objective in the survey may provide realistic view of the sustainability of farm and farming activities.

**Objective II: Identify Predictors of SAPs Adoption using Farm Attributes, Farmers' Attitudes and Behaviors, Socioeconomic and Demographic Factors and Knowledge:**

**Adoption Status of Sustainable Agricultural Practices:**

Thirty-one commonly adopted sustainable agricultural practices were listed in the survey to assess the adoption of those practices. Sustainable agricultural practices listed in the survey and their descriptions are listed in Appendix D. The response counts of all thirty-one practices are listed in the appendix E. The result shows that "Manure Distribution as Fertilizer" was the most adopted practices among listed practices followed by "Reduced Use of Chemical Pesticides" (Figure. 3), though Battle and Kruger (2005) found that use of manure was not preferred among Missouri farmers.

The result suggests that farmers have adopted practices that can be easily adopted in their farm. Respondents of the survey were more interested to reduce the use of agricultural inputs that have negative impacts in environment and interested to adopt technologies that have positive impacts in environment. Also, practices such as Manure Distribution, Cover Cropping and Green Manuring, Crop Rotation are important to improve soil quality and biological density in soil (Gabrielyan *et al.,* 2010; Carlisle, 2016). These practices are not only directly associated with improving the quality of environment and soil, but also are less expensive and easily available, accessible and affordable technologies for farmers. The literature suggests that cheap and easily adoptable technologies are more likely to be adopted by farmers as well as more likely to spread quickly among farmers (Krause and Black, 1995; Lambert *et al.*, 2006).

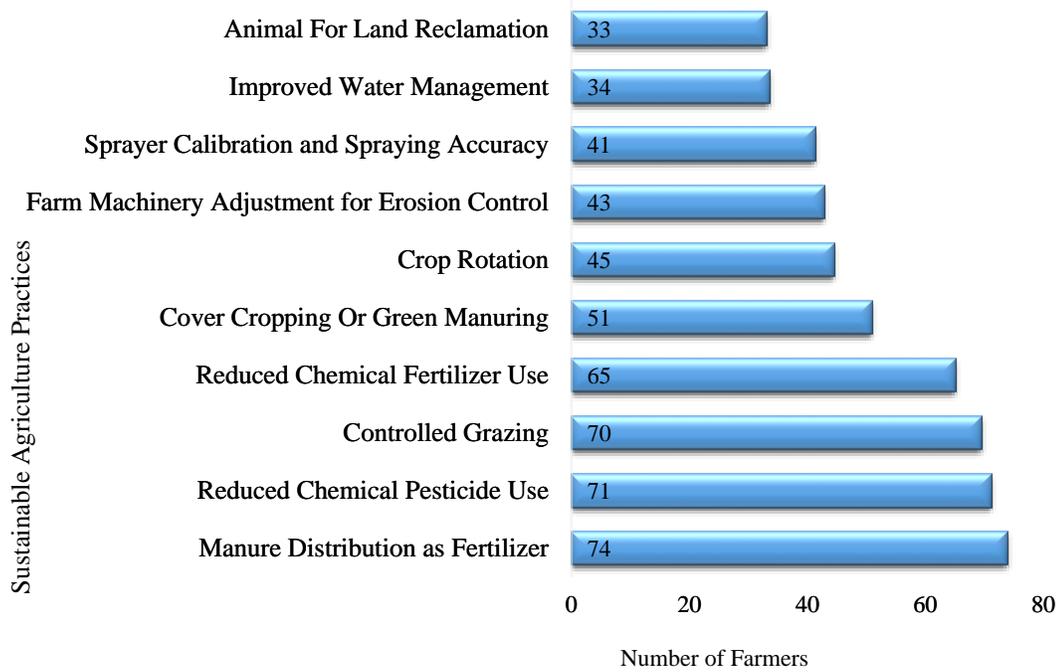

**Figure 2:** Ten most adopted sustainable agricultural practices among Kentucky farmers

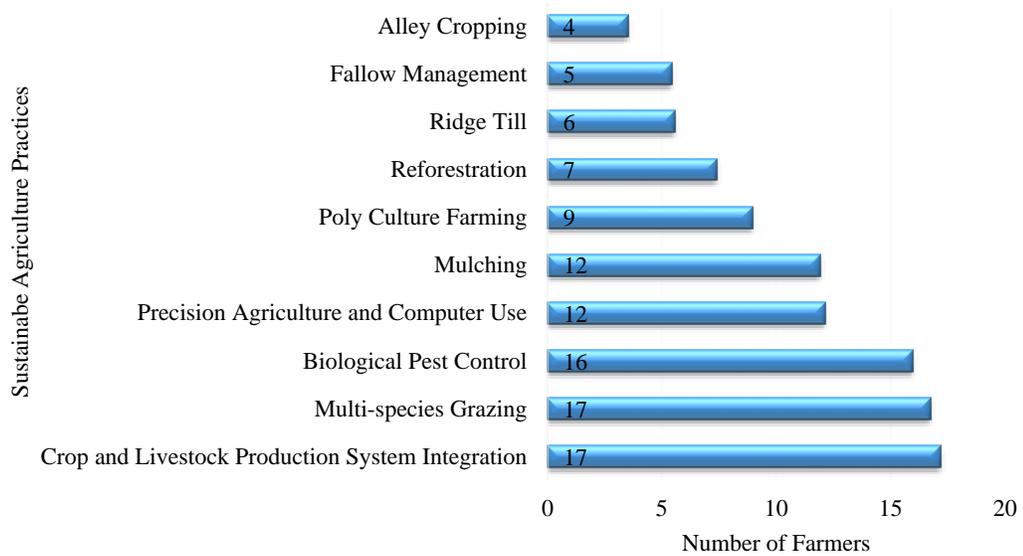

**Figure 3:** Ten Least Adopted Sustainable Agricultural Practices by Kentucky Farmers

**Regression Model Prediction of Sustainable Agricultural Practices Adoption:**

**Dependent Variable:**

**Table 3:** Total Number of Sustainable Agricultural Practices Adopted by Individual Farmers in Kentucky

| Number of Practices | Frequency | Percent | Cumulative Percent |
|---|---|---|---|
| 0 (Not Adopted) | 150 | 65.32 | 65.3 |
| 1 to 7 | 51 | 22.37 | 87.7 |
| 8 to 14 | 22 | 9.39 | 97.1 |
| 15 to 21 | 5 | 2.27 | 99.3 |
| 22 to 28 | 2 | 0.65 | 100.0 |
| Total | 230 | 100.0 | |

Number of practices adopted by farmers from 31 commonly identified practices, was counted and considered as a dependent variable in the regression model. The practices listed in the survey are provided in appendix D with its descriptions. The table shows that 65.32% of farmers were non-adopters and 22.37% and 9.39% of farmers adopted 1-7 and 8-14 practices respectively. Single farmer had adopted at most 28 practices out of 31 identified sustainable practices (Table 3).

**Explanatory Variables:**

The adoption of SAPs has been governed by various factors. Socioeconomic, demographic, farm attributes and other related variables were used to predict the relation with adoption of sustainable agricultural practices among Kentucky farmers. Ten categorical variables—Crops, Vegetables, Livestock, Tobacco Buyout Program, Diverse, Inadequate Knowledge, Irrigation, Off Farm Work, Solo Proprietorship, Income from Agri-tourism—and four continuous variables—Land Operated, Age, Education Level, Income from Farming—were used in the analysis. The description of variables is in Table 4. Descriptive statistics of the variables are presented in the Appendix C, Table 13.

**Table 4**: List of Variables in the Negative Binomial Regression and Their Descriptions:

| Variables | Variable Name | Types | Relation with DV |
|---|---|---|---|
| **SAPs** | Number of SAPs adopted by Farmers | Count | DV |
| **Crops** | Row Crop Farmers Yes = 1; Otherwise = 0 | Binary | +Ve |
| **Veggies** | Vegetable Growing Farmers Yes = 1; Otherwise = 0 | Binary | +Ve |
| **Livestock** | Livestock Farmers Yes = 1; Otherwise = 0 | Binary | -Ve |
| **TBP Participation** | Participated Yes = 1; Otherwise = 0 | Binary | +Ve |
| **Diverse** | In Favor of Diversifying Farm Yes = 1; Otherwise = 0 | Binary | +Ve |
| **Inadequate Knowledge** | A reason for not adopting SAP. Yes = 1; Otherwise = 0 | Binary | -Ve |
| **Irrigation** | Irrigation Facility in Farm Yes = 1; Otherwise = 0 | Binary | +Ve |
| **Work Off Farm** | Working off Farm Yes = 1; Otherwise = 0 | Binary | +Ve |
| **Sole Proprietorship** | Farm with Sole Proprietorship Yes = 1, Otherwise = 0 | Binary | +Ve |
| **Agro-tourism Income** | Draw income form Agro-tourism Yes = 1; Otherwise = 0 | Binary | +Ve |
| **Land Operated** | Total Land Operated (Acres) | Continuous | +Ve |
| **Age** | Age (Years) | Continuous | -Ve |
| **Education Level** | Formal Education Level | Ordinal | +Ve |
| **Income from Farming** | Income from Farming | Ordinal | +Ve |

**Negative Binomial Regression Model Specifications and Result Interpretations:**

Independent variables consist of 10 categorical variables (Crops, Veggies, Livestock, TBP Participation, Diverse, Inadequate Knowledge, Irrigation, Work Off-farm, Sole Proprietorship, and Agro-tourism Income) with 1 as "Yes" and 0 as "No" categories. For example, a farmer was asked "Do you grow crops in your farm?". If a farmer replied "Yes", then the value is entered as 1, otherwise it is 0; similar coding method was used for all other binomial variables. The variable "Sole Proprietorship", farmers who own the farm solely was dummy coded as 1 and other responses such as—partnership, family held corporation, trust or estate, and others or missing—were coded as 0. The code 1 was used as the reference category for binomial variables in the model.

Two continuous variables Land Operated (Acres) and Age of farmers were used in the model. "Education Level" and "Income" (from farming) were categorical variables used as covariates in the model. The variable "Education Level" has six categories—below high school, high school degree, some college degree, college degree, and graduate or professional degree. The lowest education level, below high school, was coded as 1 the graduate or professional degree as 5. The variable "Income from Farming" also has six categories—below $10,000, $10,000 to $49,999, $50,000 to $99,999, $100,000 to $249,999, $250,000 to $500,000 and more than $500,000. Smallest category is coded is 1 and largest as 6.

The regression model analyzed 205 cases out of 230 respondents. Ten cases with more than 3,000 acres of land operated were removed from the model considering them as potential outliers in the variable. The status of respondents, though, has been described based on 230 cases earlier, the descriptive statistics of 205 cases used in the model is also included in the appendix section along with the results of all four models Appendix F.

The Goodness-of-Fit of value (1.972) indicates over-dispersion. The Omnibus Test is significant with Likelihood Ratio $\chi^2$ (13) = 93.142, $p$ = 0.000, which shows that the model is significant over the null model—a model without predictors variables or intercept-only model (Coxe *et al.*, 2009).

A Negative Binomial Regression model (due to over-dispersion) was used to predict the SAP adoption of Kentucky farmers. The regression analysis shows that Crops, Veggies, Diverse, Irrigation, and Education are significant variables at 95% CI. Agro-tourism Income, Land Operated, Age are significant at 90% CI but, Livestock, TBP, Inadequate Knowledge, Off Farm-work, Sole Proprietorship are not significant. Baumgart-Getz *et al.* (2012) found knowledge as significant predictor of adoption, however, the variable related to knowledge was not significant in this model. All variables except sole proprietorship in the model retained predicted signs. The hypothesis was accepted for all variables except sole proprietorship in regards to the direction of relationship.

**Table 5:** Parameter Estimates of the Negative Binomial Regression and Goodness-of-Fit Table.

| VARIABLES | *B* | STD. ERROR | WALD CHI-SQUARE | SIG. | EXP. (*B*) |
|---|---|---|---|---|---|
| CONSTANT | -0.305 | 0.9138 | 0.111 | 0.739 | 0.737 |
| CROPS*** | 1.067 | 0.2388 | 19.967 | **0.000** | 2.907 |
| VEGGIES** | 0.555 | 0.2816 | 3.892 | **0.049** | 1.743 |
| LIVESTOCK | 0.363 | 0.3386 | 1.147 | 0.284 | 1.437 |
| TBP | 0.181 | 0.2152 | 0.705 | 0.401 | 1.198 |
| DIVERSE*** | 0.778 | 0.2100 | 13.730 | **0.000** | 2.178 |
| INADEQUATE KNOWLEDGE | -0.254 | 0.2706 | 0.882 | 0.348 | 0.776 |
| IRRIGATION** | 0.948 | 0.4261 | 4.954 | **0.026** | 2.582 |
| WORK OFF FARM | 0.109 | 0.2632 | 0.172 | 0.679 | 1.115 |
| SOLE PROPRIETORSHIP | -0.070 | 0.2717 | 0.067 | 0.796 | 0.932 |
| AGRO-TOURISM INCOME* | 0.927 | 0.5187 | 3.193 | **0.074** | 2.526 |
| LAND OPERATED* | 0.001 | 0.0004 | 3.620 | **0.057** | 1.001 |
| AGE* | -0.022 | 0.0118 | 3.611 | **0.057** | 0.978 |
| EDUCATION LEVEL*** | 0.305 | 0.0822 | 13.771 | **0.000** | 1.357 |
| *** | 99% Confidence Interval (0.001 Significance Level) | | | | |
| ** | 95% Confidence Interval (0.05 Significance Level) | | | | |
| * | 90% Confidence Interval (0.1 Significance Level) | | | | |

| GOODNESS OF FIT | VALUE | DF | VALUE/DF |
|---|---|---|---|
| DEVIANCE | 361.122 | 191 | 1.891 |
| SCALED DEVIANCE | 361.122 | 191 | |
| PEARSON CHI-SQUARED | 376.730 | 191 | 1.972 |
| SCALED PEARSON CHI-SQUARED | 376.730 | 191 | |
| LOG LIKELIHOOD | -404.668 | | |
| AKAIKE'S INFORMATION CRITERION (AIC) | 837.337 | | |
| FINITE SAMPLE CORRECTED AIC (AICC) | 839.547 | | |
| BAYESIAN INFORMATION CRITERION (BIC) | 88.3859 | | |
| CONSISTENT AIC (CAIC) | 897.859 | | |
| N | 205 | | |

*Farmers cultivating row crops* (Crops) variable is significant *(p = 0.000)* and has a positive relationship with the adoption of SAPs in Kentucky. The change in the expected log count of SAPs adoption would be expected to increase by 1.067 if farmers have row crops in their farm. Several researchers have found that farmers diversify their farms by adding high value crops, short season crops and cover crops (Lambert *et al.*, 2006; Lichtenberg 2004; Singer *et al.*, 2007; Snapp *et al.*, 2005). Cover crops are helpful in contributing weed control, reduce use of chemicals, soil health, fertility, and improved crop performance thus sustaining the production (Teasdale, 2013).

*Farmers growing vegetables in their farm (veggies)* variable is positively significant *(p =0.049)* and has positive correlation with the SAP adoption behavior in Kentucky. The change in the expected log count of SAPs adoption would be expected to increase by 0.555 if farmers have vegetables in their farm other things remaining constant. Tey *et al.* (2013) found that vegetable farmers in Malaysia are highly influenced by conservation practices and IPM.

*Farmer's willingness to diversify their farm* (diverse) is significant *(p = 0.000)* and positive. The change in the expected log count of SAPs adoption would be expected to increase by 0.778 if farmers are in favor of diversifying their farm. Baumgart-Getz *et al.* (2012) found that the positive attitudes towards farm diversification has positive and significant influence in the adoption of agricultural best management practices.

*Farmers having farm with the irrigation facilities* (irrigation) is another positive and significant variable *(p = 0.026)*. The change in the expected log count of SAPs adoption would be expected to increase by 0.948 if farmers have irrigation facilities in their farm. The finding of the research is consistent with Alabama cover crop farmers (Bergtold *et al.*, 2012).

Education variable is also significant and positive *(p = 0.000)* predictor of the adoption of sustainable agricultural practices among Kentucky farmers. The change in the expected log count of SAPs adoption would be expected to increase by 0.305 with an increase in the education level. However, the coefficient may vary according to the level of education as the change in knowledge or education between for instance, high school and college may not be equivalent to that between college and graduate degree. Therefore, this variable needs to be reclassified in the model to see the effect of different level of education in the adoption of SAPs among Kentucky farmers.

Baumgart-Getz *et al.* (2012) found that level of education was non-significant among large number of farmers but extension training had positive impact on farmer's adoption behavior among US farmers. In another study by Hall *et al.* (2009) the education level was not significant to the adoption of SAPs among floriculture farmers in Midwest USA. Baumgart-Getz *et al.* (2012) also found that overall education category and formal education were non-significant, but extension training was significant and had positive impact on the adoption behavior of farmer.

The "Age" of farmer *(p = 0.057)* and "Land Operated" *(p = 0.057)* are also significant variables (at 90% CI and close to 95% CI) and "Income from Agro-tourism" *(p = 0.074)* (at 90% CI) of SAPs adoption behavior of Kentucky farmers, but their direction of effect varies. Income from agro-tourism and land operated have positive correlations with the adoption of SAP whereas the Age has a negative one. The change in the expected log count of SAPs adoption would be expected to increase by 0.927, if farmers have agro-tourism as a source of farm income, other things remaining constant. However, the age has negative effect in the adoption decision. For each one unit (year) increase on age of farmers, the expected log count of SAPs adoption decrease by 0.022.

The Age does not have a consistent predictability in several researches. Age was found to be a non-significant predictors of SAPs adoption among floriculture farmers in US (Hall *et al*., 2009) but reported significant by Kabi & Horwitz (2006) suggesting age is a significant predictor of the easement towards the adoption of conservation practices. Baumgart-Getz *et al.* (2012) also found that age as a significant predictor of adoption of best management practices in agriculture with negative effect. Prokopy *et al.* (2008) found that, more often, age has negative relation with the adoption rather than positive and never has positive relation with nutrient or water management related practices. However, old beef cattle farmers are more likely to adopt soil and landscape management than younger farmers. This could be because older beef cattle farmers are considering farming as hobby and place high importance in land (Prokopy *et al.*, 2008).

The variable land operated *(p = 0.057)* appeared to be a significant variable at 90% CI and close to 95% CI. For each one-unit acres of increase on land operated the expected log count of adoption of SAPs increase by 0.001, other things remaining constant. Hall *et al.* (2009) also found that land operated is a significant predictor, with greater odds, of adoption behavior among Midwest small scale floriculture farmers but size alone cannot predict the adoption.

**Negative Binomial Regression Models with Outliers in the Variable Land Operated and the Variable Income:**

The negative binomial regression with log link function was run with other three combinations of the models and variables: 1) with the annual income from farming (income) and with outliers—farmers having above 3,000 acres of operated land—in the variable land operated, 2) without income and with outliers in the land operated and 3) with income and without outliers in the land operated. The variables whose level of significance has changed in those models are discussed in this section in the order models are presented in the beginning of this paragraph. The summary and model details are provided in Appendix F.

The variable "Veggies" became non-significant at 95% CI but significant at 90% CI in all other models. In the third model it came close to 95% CI with *(p = 0.056)*.

Income from agro-tourism became significant at 95% CI in the first *(p = 0.029)* and the second *(p = 0.057)* models. The variable Land Operated became non-significant in the first and the second models and remained significant at 90% CI *(p = 0.087)* in the third model, however, the coefficients remained virtually constant.

**Objective III: Adoption Barriers of Sustainable Agricultural Practices:**

The survey was designed to identify the potential barriers of the adoption of sustainable agricultural practices. The survey composed of close end questions with choices provided to mark, if any of the barriers listed were barriers for the adoption of sustainable agricultural practices. Also, "Others" option was provided with space to write if farmers had any other barriers other than listed in the survey.

"Inadequate Knowledge" (reported by 15.22% respondents) was found to be the major adoption barrier of SAPs followed by "Perceived Difficulty of Implementation" (5.22% of respondents). Inadequate Knowledge was not significant in the regression model. However, it was reported by a moderate proportion of respondents. Moreover, "Lack of Adequate Market" and "Consumer Preference for Alternative Products" were reported as the barriers by 3.48% of the respondents. These two reasons could be closely related as market and consumer preference are interrelated.

**Table 6:** Adoption Barriers of SAPs among Kentucky Farmers

| Listed Adoption Barriers in Survey | Frequency | Percent |
|---|---|---|
| Inadequate Knowledge | 35 | 15.22 |
| Perceived Difficulty of Implementation | 12 | 5.22 |
| Negative Attitude About Technologies | 8 | 3.48 |
| Lack of Adequate Market for Alternative Product | 8 | 3.48 |
| Lack of Appropriate Technologies | 7 | 3.04 |
| Lack of Consumer Acceptance for Alternative Product | 3 | 1.30 |
| Happy With What I am Doing | 99 | 43.04 |
| Others | 7 | 3.04 |
| Total | 183 | 79.57 |
| Missing | 47 | 20.43 |
| Grand Total | 230 | 100.00 |

The result also shows that large proportion (43.04%)of respondents have "Happy" attitude about their farm and agricultural practice they have adopted. However, the reason behind the happiness of farmers about their farm and
farming practices is not well understood, which could be
their satisfaction to anything they have been practicing
from the perspective of income, production, profit,
environment health, quality of life, greenery or farmer's
unwillingness to adopt new practices in their farm due to
high risk associated with changing pattern. Others options
in the survey have reasons that, one way other, are closely
related with the reasons listed in the survey. The
frequency distribution and the percentage of adoption
barriers of SAPs is provided in Table 6 above.

Similar findings have been supported by various
researches. Baumgart-Getz *et al.* (2012) mentioned that
variables related to awareness such as education, causes,
consequences, knowledge, and programs are important to form
an attitude towards the adoption of management practices.
The profitability of practices has positive impact in the
adoption process. Charlisle (2016) found that farmers'
knowledge and access to information are important factors
in influencing the adoption of practices related to soil
health. Farmers with higher level of knowledge about benefits of certain practices over others are more likely to adopt these practices and technologies compared to farmers without or less knowledge.

The "Happy" attitude of farmers may be somehow associated with the age factor. The average age of the respondents in this research was high. Baumgart-Getz *et al.* (2012) and Prokopy *et al.* (2008) found that age had negative significant relation with the of adoption of best management practices. Older age farmers considered farming as hobby and placed high importance to the possession of land. However, it is subject to research in future to develop better understanding of the adoption barriers and adoption of sustainable agricultural practices.

**CHAPTER V**

**CONCLUSIONS AND RECOMMENDATIONS**

Sustainability in agriculture is a core of research topic for decades. It is dynamic-changing over time, location, community and people.  Sustainability has been defined from multidimensional perspectives such as managerial, philosophical and system of farming—and a holistic—integrating socioeconomic, cultural, financial and environmental perspectives. This thesis explored farm level goals of sustainable practices from farmers' perspectives and agriculture sustainability from the managerial perspectives.

In regards to Objective I, the question about farmers' perception of achieving sustainability goals majority of the responding farmers rated above 3 in the rating scale of 1 to 5, which indicates that farmers, overall, are satisfied with the sustainability of their farm and farming practices. Farmers perceive that they meet the objectives of agriculture sustainability. However, agriculture sustainability is a dynamic term and requires to study additional measures to properly measure the objectives of sustainability. The variables associated with the objective
I might be broad and may have not been connected to their
sustainability practices. The perception of sustainability
differs from each other, usually sustainability is
understood as environmental sustainability, although, it is
a holistic concept (Allen *et al.*, 2013). Therefore, the
results should be understood as more general perception of
farmers about achieving sustainability in their farm.
However, this is an important finding to explore
continuously with more variables that measure the
sustainability objective in very precise form.

Adoption of SAPs assists to attain agriculture
sustainability. The research identified some of the major
predictors of the adoption of SAPs in Kentucky. The results
suggest that crop and vegetable growing farmers are
significant and positive predictors of the adoption of
sustainable agricultural practices, but livestock farmers
are not though they indicate positive relation with the
adoption. However, vegetable growing farmers turned to be
non-significant predictor when outliers of the variable
"land operated" was added into the model. Irrigation and Education are also two other positive significant predictors of the SAPs adoption. Farmers generating income from agro-tourism variable is also significant and positive predictor of the adoption behavior at 90% confidence interval. However, when income and outliers in the variable "land operated" were added in the model, the variable became significant at 95% confidence interval. The age has negative but significant relation with the adoption of sustainable agricultural practices. Also, the variable land operated became significant in the model with income and without outliers at 90% confidence interval which was non-significant in other models. But the effect of land operated is virtually null in the model.

The model is also significant compared to the constant-only model and also had several significant predictors of the adoption behavior. However, the dependent variable has more than 50% 0s or not adopters of SAPs and also was overly dispersed. The negative binomial regression model addressed the problem of over-dispersion, however, a comparison of existing model with zero inflated negative binomial regression model using Poisson-quasi likelihood parameters, non-parametric or semi-parametric models could be useful to predict significance level and intensity of the variables in the adoption behavior more precisely addressing excess not adopters of SAPs (0s) in the model.

Inadequate Knowledge and Implementation difficulties were identified as two major barriers for the adoption of SAPs. The implementation difficulties also arise, completely or partially, from inadequate knowledge. This emphasize the importance of extension programs to promote adoption of sustainable agricultural practices among farmers.

The adoption process can be improved significantly by increasing environmental awareness, networking capacity of farmers (with both agencies and local networks), and information about practices; they have significant impact in the adoption of those practices. The use of technologies can be further improved by increasing extension service, diversifying farming tools, external support from government and non-governmental bodies (Barungi *et al.*, 2013). Farmers motivated to adopt sustainable agricultural practices by making farmers knowledgeable about action of individual affect their own life quality rather than stating how the life is degraded by not adopting environment friendly management practices in their farm and implementation of program that helps to reduce the impact in their farm level rather than in general to overall environment (Baumgart-Getz *et al.*, 2012).

Another important reasons to consider while developing extension programs are "Happy" attitude of farmers. It is very important to understand the sources of happiness among farmers as it is mentioned by almost half (*43.04%*) of the farmers. The happiness could be from anywhere: farmers' satisfaction from yield and income, farmers' satisfaction as they are able to utilize their retired life making some extra money and keeping themselves active in their farm, their contribution to environment conservation, their affinity with crops, vegetables and livestock, hobbies or their unwillingness to maximize their production and profit, or unresponsive nature.

This study possesses some limitations which should be taken into consideration while interpreting results adopting them for education and extension. Since majority of respondents are White farmers followed by African American farmers, the implication of policies might be limited. Also, participating farmers are small farmers and thus findings are more relevant to the small farmers. The gender related information were not collected in the research which makes research implacable to the overall farmers in general.

In addition, the stratification has reduced the sampling bias in the research. However, it is important to analyze the data including spatial components into the analysis as the characteristics of agriculture varies along the state. Eastern Kentucky has less agriculture land compared to western Kentucky which might have impact in the decision making process in agriculture and also have effect in the findings of research.

## Appendix A: Survey Distribution in Each Strata

**Table 7**: Distribution of Survey Dispactched in Each Strata. This table shows total number of survey sent in all 24 strata throughout the state. Agriculture district map and coding is provided in the appendices section of the report.

| Ag District | Annual Value of Sales (Annual Income) | | | | Grand Total |
|---|---|---|---|---|---|
| | 1 | 2 | 3 | 4 | |
| 10 | 67 | 45 | 40 | 26 | 178 |
| 20 | 67 | 50 | 41 | 29 | 187 |
| 30 | 66 | 46 | 37 | 23 | 172 |
| 40 | 66 | 43 | 28 | 4 | 141 |
| 50 | 66 | 50 | 40 | 14 | 170 |
| 60 | 65 | 49 | 34 | 4 | 152 |
| Grand Total | 397 | 283 | 220 | 100 | 1000 |

# Appendix B. Stratification Variable and Categorization:

**Table 8:** Four Strata Developed based on Approximate Value of Sales of Kentucky Farmers.

| Strata Min | Strata Max | Strata |
|------------|------------|--------|
| 0 | $60,000 | 1 |
| $60,000 | $350,000 | 2 |
| $350,000 | $2,000,000 | 3 |
| 2,000,000 | $50,000,000 | 4 |

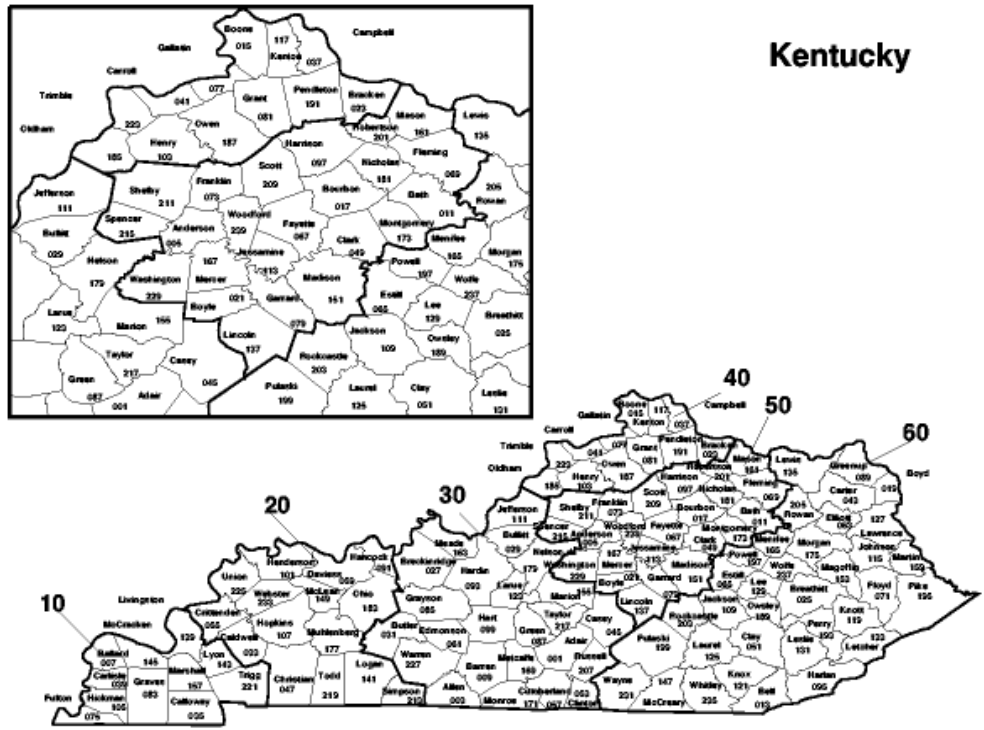

**Figure 4:** Agriculture District of Kentucky. It is numbered as 10, 20, 30, 40, 50 and 60 starting from Western Kentucky to Eastern Kentucky moving towards North.

**Appendix C: Frequency Distribution of Demographic, Socioeconomic and Farm Attribute Related Variables in the Survey.**

**Table 9:** Annual Income and Ownership of Farm land among Kentucky Farmers

| Income | Frequency | Percent | Ownership | Frequency | Percent |
|---|---|---|---|---|---|
| Below 10K | 95 | 41.3 | Sole Proprietorship | 172 | 74.7 |
| 10K to <50K | 63 | 27.2 | Partnership | 13 | 5.8 |
| 50K to <100K | 13 | 5.6 | Family Held Corporation | 7 | 3.1 |
| 100K to <250K | 9 | 3.9 | Corporation (Not Family Held) | 0 | 0.1 |
| 250K to <500K | 5 | 2.0 | Trust or Estate | 7 | 3.0 |
| Above 500K | 3 | 1.3 | Others | 3 | 1.3 |
| Total | 187 | 81.3 | Total | 202 | 87.9 |
| Missing | 43 | 18.7 | Missing | 28 | 12.1 |
| Total | 230 | 100.0 | Total | 230 | 100.0 |
| **Note:** K = $1,000 | | | | | |

**Table 10**: Frequency Distribution of Age, In Farming and Decision Making (Years) of Kentucky Farmers

| Years Group | Age | | In Farming | | Decision Making | |
|---|---|---|---|---|---|---|
| | Frequency | Percent | Frequency | Percent | Frequency | Percent |
| 0 to 10 | 0 | 0 | 25 | 11.0 | 31 | 13.7 |
| 11 to 20 | 0 | 0 | 32 | 13.8 | 34 | 14.8 |
| 21 to 30 | 0 | 0.1 | 39 | 16.9 | 33 | 14.5 |
| 31 to 40 | 6 | 2.4 | 19 | 8.4 | 30 | 13.2 |
| 41 to 50 | 36 | 15.8 | 40 | 17.5 | 39 | 16.8 |
| 51 to 60 | 41 | 17.9 | 32 | 13.8 | 17 | 7.6 |
| 61 to 70 | 59 | 25.8 | | | 0 | 0 |
| 71 and Above | 65 | 28.1 | | | 0 | 0 |
| Total | 207 | 90.1 | 187 | 81.4 | 185 | 80.7 |
| Missing | 23 | 9.9 | 43 | 18.6 | 45 | 19.3 |
| Total | 230 | 100.0 | 230 | 100.0 | 230 | 100.0 |

**Table 11**: Formal Education Level of Kentucky Farmer and Their Spouse

| Education Level | Farmer | | Farmers' Spouse | |
|---|---|---|---|---|
| | Frequency | Percent | Frequency | Percent |
| Less than High School | 30 | 13.0 | 22 | 9.6 |
| High School | 71 | 30.8 | 47 | 20.5 |
| Some college | 44 | 19.1 | 17 | 7.3 |
| College Degree | 26 | 11.2 | 23 | 10.0 |
| Graduate Degree | 21 | 9.0 | 33 | 14.5 |
| Total | 191 | 83.1 | 142 | 61.9 |
| Missing | 39 | 16.9 | 88 | 38.1 |
| Total | 230 | 100.0 | 230 | 100.0 |

**Table 12:** Total Land Operated, Owned and Rented (Acres) by Kentucky Farmers

| Acres | Land Operated | | Land Owned | | Land Rented | |
|---|---|---|---|---|---|---|
| | Frequency | Percent | Frequency | Percent | Frequency | Percent |
| 0 to 99 | 116 | 50.3 | 123 | 53.5 | 32 | 14.1 |
| 100 to 249 | 77 | 33.5 | 69 | 30.2 | 14 | 6.2 |
| 250 to 499 | 25 | 11.0 | 21 | 9.2 | 7 | 3.0 |
| 500 to 999 | 8 | 3.6 | 10 | 4.6 | 4 | 1.7 |
| 1000 to 1999 | 4 | 1.5 | | | | |
| 2000 to 2999 | | | | | | |
| 3000 and Above | | | | | | |
| Total | 230 | 100.0 | 224 | 97.4 | 57 | 25.0 |
| Missing | 0 | 0.0 | 6 | 2.6 | 173 | 75.0 |
| Total | 230 | 100.0 | 230 | 100.0 | 230 | 100.0 |

**Table 13:** Descriptive Statistics of Variables

| Variable | N | Mean | Std. Error of Mean | Std. Deviation | Variance |
|---|---|---|---|---|---|
| Crops | 151 | 0.543 | 0.019 | 0.346 | 0.154 |
| Veggies | 230 | 0.1597 | 0.02421 | 0.36710 | 0.135 |
| Livestock | 221 | 0.81 | 0.027 | 0.396 | 0.157 |
| TBP Participation | 191 | 0.42 | 0.045 | 0.617 | 0.380 |
| Diverse | 204 | 0.42 | 0.035 | 0.495 | 0.245 |
| Inadequate Knowledge | 230 | 0.15 | 0.024 | 0.361 | 0.130 |
| Irrigation | 230 | 0.0378 | 0.01261 | 0.19120 | 0.037 |
| Work Off Farm | 210 | 0.49 | 0.035 | 0.501 | 0.251 |
| Sole Proprietorship | 230 | 0.75 | 0.029 | 0.436 | 0.190 |
| Agro-Tourism Income | 193 | 0.03 | 0.013 | 0.182 | 0.033 |
| Land Operated | 230 | 169.596 | 36.165 | 548.457 | 300804.589 |
| Land Owned | 224 | 144.597 | 22.275 | 333.445 | 111185.604 |
| Land Rented | 57 | 197.582 | 94.545 | 716.441 | 513287.755 |
| Age (Years) | 207 | 62.85 | 0.850 | 12.240 | 149.813 |
| In Farming (Years) | 187 | 33.28 | 1.352 | 18.487 | 341.781 |
| Decision Making (Years) | 185 | 29.86 | 1.200 | 16.337 | 266.901 |
| Farmer's Education Level | 191 | 2.67 | 0.088 | 1.210 | 1.464 |
| Spouse Education Level | 142 | 2.99 | 0.120 | 1.437 | 2.064 |
| Income | 187 | 1.80 | 0.081 | 1.101 | 1.212 |

**Appendix D: Terms and Definition of Commonly Adopted Sustainable Agricultural Practices among Kentucky Farmers.**

Commonly adopted SAPs used in this research are defined as follows:

*Alley Cropping:* It is the practices of planting trees or shrubs in two or more sets of single or multiple rows with agronomic, horticultural or forage crops cultivated in the alleys between the rows of woody plants (Kornegay *et al.,* 2010).

*Animal for Land Reclamation:* Excessive compaction of soil results in delayed ecological succession and low species diversities on reclaimed mine lands. Small mammals such as mouse help to loosen the mined surface which favors quick succession (Larkin *et al.,* 2008).

*Biological Pest Control:* Pest are suppressed by their natural enemies, such as birds, spiders, insects, miter, fungi, bacteria, viruses or plants (Filho *et al.,* 1999).

*Composting:* It is an important organic waste (from post harvesting processing) recycling technique which is converted into humus due to breakdown by micro-organisms and soil fauna.

It improves quality and quantity of soil organic matters (Filho *et al.*, 1999).

**Conservation Tillage:** Set of practices that level crop residues on the surface which increases water infiltration and reduce soil erosion (Hobbs, Sayre, & Gupta, 2008).

**Controlled Grazing:** The degree of control or level of management applied to grazing animals, through the use of such grazing systems as rotational stocking, continuous stocking and strip grazing (White & Wolf, 2009).

**Cover Crops and Green Manuring:** Cover crops and green manures are grown for the benefit they provide to the production of subsequent cash crops. Cover crops protects the bare soil against erosion and green manures improves soil fertility. As both cover crops are also added to the soil and coverts into green manure, these two terms are used interchangeably (Kornegay *et al.*, 2010).

**Crop and Livestock Diversification:** Diversification of crops or livestock. The crop and livestock production systems may or may not be integrated as a mixed farming system (Kornegay *et al.*, 2010).

***Crop and Livestock Production System Integration:*** An integrated system where crop and livestock enterprise are combined and benefitted from each other (Kornegay *et al.*, 2010).

***Crop Rotation:*** System of cultivation where different crops are planted in consecutive growing seasons to maintain soil fertility (Kornegay *et al.*, 2010).

***Cultural Pest Control:*** Managing the crop, weed, disease and pest complex by manipulating cultural practices (Kornegay *et al.*, 2010).

***Fallow Management:*** The use of fallow period to conserve rainfall as stored soil water and reduce soil erosion (Kornegay *et al.*, 2010).

***Farm Machinery Adjustment (for Erosion Control):*** Adjustment in planting, spraying and harvesting farm machinery operation, calibration, repair, and their safety (Kornegay *et al.*, 2010).

***Forest Stewardship:*** Forest conservation and development of forest in own farm land, use of forest products from well-managed forest and use of recycled materials.

***Improved Water Management:*** Use of improved irrigation facilities to reduce losses of irrigation water (Kornegay *et al.*, 2010).

***Increase Biodiversity:*** Increase diversity by adding various plants and animals in the farm (Kornegay *et al.*, 2010).

***Integrated Pest Management:*** A pest management strategy using a systematic approach in which pest population are monitored to determine if and when control methods are required. It uses biological, chemical, physical, cultural production cost and protect the environment (Kornegay *et al.*, 2010).

***Land Reforming (to reduce erosion):*** Forming terrace, reducing slope, and other slope stabilizing technologies to reduce surface run off of water and top soil.

***Local or Native Crops:*** Locally available crops or local varieties (Kornegay *et al.*, 2010).

***Mulching:*** A shallow layer at the soil/air interface formed by dry grass, crop residuals, fresh organic materials, weeds, cover crops and green manures to improve soil microclimate and enhance soil life, structure, fertility, moisture, suppress weeds, reduce solar and rainfall damage (Filho *et al.*, 1999).

***Multi-species Grazing:*** Grazing more than one species of types of livestock together in same land. Egs. Chicken, duck and goat grazing together (Kornegay *et al.*, 2010).

***Poly-culture Farming:*** Different species can be used on lands with different characteristics to optimize biomass yield and improve environmental quality (Kornegay *et al.*, 2010).

***Precision Agriculture (and Computer Use):*** Farming management concept based on observing, measuring and responding to inter and intra-field variability in crops and helps to increase farm's productivity, input-use efficiency, and economic returns (Kornegay *et al.*, 2010).

***Reduced Chemical Fertilizer Use:*** Reduced in the use of chemical fertilizers (Kornegay *et al.*, 2010).

***Reduced Chemical Pesticide Use:*** Reduce in the use of chemical pesticides (Kornegay *et al.*, 2010).

***Reforestation:*** Reestablishing forest in barren land or farm land.

**Ridge Tillage:** A tillage system involving scalping and planting on ridges built during cultivation of the previous year's crop, usually involves spring-planted row crops grown with a combination of herbicides and at least one cultivation (Kornegay *et al.*, 2010).

**Sprayer Calibration (and Application Accuracy):** Calibrate sprayers to use optimum amount of chemicals as well as other spraying inputs in farm.

**Varietal Mixture of Single Crop:** Mixing different variety of same crops. Also known as Cultivar Mixtures (Kornegay *et al.*, 2010).

**Windbreaks and Shelterbelts:** Environment buffers that are planted in a variety of settings, such as on cropland, pasture, rangeland, along road, farmstead, feedlots and in urban areas to reduce negative consequences of winds in crops or sites (Kornegay *et al.*, 2010).

**Table 14:** The Adoption Status of Listed Sustainable Agricultural Practices by Kentucky Farmers

| Sustainable Agricultural Practices | Yes | No | Planning | N/A | Total |
|---|---|---|---|---|---|
| Manure Distribution as Fertilizer | 74 | 31 | 2 | 33 | 140 |
| Reduced Chemical Pesticide Use | 71 | 47 | 7 | 33 | 158 |
| Controlled Grazing | 70 | 36 | 1 | 36 | 143 |
| Reduced Chemical Fertilizer Use | 65 | 45 | 9 | 31 | 149 |
| Cover Cropping and Green Manuring | 51 | 41 | 5 | 43 | 140 |
| Crop Rotation | 45 | 32 | 0 | 59 | 136 |
| Farm Machinery Adjustment(for Erosion Control) | 43 | 48 | 0 | 51 | 141 |
| Sprayer Calibration (and Spraying Accuracy) | 41 | 48 | 0 | 52 | 142 |
| Improved Water Management | 34 | 54 | 0 | 57 | 145 |
| Animal For Land Reclamation | 33 | 58 | 3 | 51 | 145 |
| Composting | 30 | 67 | 0 | 42 | 139 |
| Land Reform (to Reduce Erosion) | 30 | 48 | 2 | 75 | 155 |
| Conservation Tillage | 30 | 47 | 0 | 74 | 151 |
| Crop and Livestock Diversification | 26 | 52 | 0 | 61 | 139 |
| Increasing Biological Diversity | 24 | 47 | 0 | 73 | 144 |
| Forest Stewardship | 23 | 29 | 6 | 83 | 141 |
| Local or Native Crops | 21 | 49 | 3 | 65 | 138 |
| Varietal Mixture of Same Crops | 21 | 45 | 0 | 71 | 137 |
| Integrated Pest Management | 19 | 63 | 0 | 70 | 152 |
| Cultural Pest Control | 19 | 58 | 0 | 61 | 138 |
| Windbreaks and Shelterbelts | 18 | 67 | 0 | 55 | 140 |
| Integration of Crop and Livestock Production System | 17 | 51 | 2 | 66 | 137 |
| Multi-species Grazing | 17 | 50 | 0 | 66 | 132 |
| Biological Pest Control | 16 | 57 | 2 | 58 | 133 |
| Precision Agriculture (and Computer Use) | 12 | 73 | 0 | 55 | 141 |
| Mulching | 12 | 71 | 0 | 55 | 138 |

| | Yes | No | Planning | N/A | Total |
|---|---|---|---|---|---|
| **Poly Culture Farming** | 9 | 47 | 3 | 79 | 139 |
| **Reforestation** | 7 | 50 | 1 | 92 | 150 |
| **Ridge Till** | 6 | 59 | 0 | 75 | 139 |
| **Fallow Management** | 5 | 60 | 0 | 75 | 141 |
| **Alley Cropping** | 4 | 57 | 2 | 76 | 139 |
| Notes: Yes = Adopted; No = Not Adopted; Planning = Planning to adopt in near future; N/A = Not Applicable for the farmer and Total = Yes + No + Planning + N/A | | | | | |

# Appendix F: Negative Binomial Regression Model Detail:

**Table 15: Model I:** NB Regression Model with Income and with Outliers in Land Operated.

| VARIABLES | *B* | STD. ERROR | WALD CHI-SQUARE | SIG. | EXP. (*B*) |
|---|---|---|---|---|---|
| CONSTANT | -0.234 | 0.9142 | 0.066 | 0.798 | 0.791 |
| CROPS*** | 1.069 | 0.2444 | 19.140 | **0.000** | 2.913 |
| VEGGIES* | 0.517 | 0.2839 | 3.315 | 0.069 | 1.677 |
| LIVESTOCK | 0.320 | 0.3362 | 0.908 | 0.341 | 1.378 |
| TBP | 0.180 | 0.2158 | 0.699 | 0.403 | 1.198 |
| DIVERSE*** | 0.759 | 0.2096 | 13.113 | **0.000** | 2.136 |
| INADEQUATE KNOWLEDGE | -0.240 | 0.2699 | 0.792 | 0.373 | 0.786 |
| IRRIGATION** | 0.946 | 0.4232 | 4.997 | **0.025** | 2.575 |
| WORK OFF FARM | 0.066 | 0.2640 | 0.063 | 0.802 | 1.069 |
| SOLE PROPRIETORSHIP | -0.038 | 0.2690 | 0.020 | 0.888 | 0.963 |
| AGRO-TOURISM INCOME** | 1.110 | 0.5093 | 4.748 | **0.029** | 3.034 |
| LAND OPERATED | 0.000 | 0.0003 | 0.410 | 0.522 | 1.000 |
| AGE* | -0.022 | 0.0117 | 3.667 | **0.056** | 0.978 |
| EDUCATION LEVEL*** | 0.291 | 0.0842 | 11.960 | **0.000** | 1.338 |
| INCOME FROM FARMING | 0.053 | 0.1034 | 0.259 | 0.611 | 1.054 |
| *** | 99% Confidence Interval (0.001 Significance Level) | | | | |
| ** | 95% Confidence Interval (0.05 Significance Level) | | | | |
| * | 90% Confidence Interval (0.1 Significance Level) | | | | |

| GOODNESS OF FIT | VALUE | DF | VALUE/DF |
|---|---|---|---|
| DEVIANCE | 365.253 | 200 | 1.826 |
| SCALED DEVIANCE | 365.253 | 200 | |
| PEARSON CHI-SQUARED | 384.937 | 200 | 1.925 |
| SCALED PEARSON CHI-SQUARED | 384.937 | 200 | |
| LOG LIKELIHOOD | -407.466 | | |
| AKAIKE'S INFORMATION CRITERION (AIC) | 844.932 | | |
| FINITE SAMPLE CORRECTED AIC (AICC) | 847.344 | | |
| BAYESIAN INFORMATION CRITERION (BIC) | 895.492 | | |
| CONSISTENT AIC (CAIC) | 910.492 | | |
| N | 215 | | |
| | | | |

| Model I: Continuous Variable Information | | N | Mean | Std. Deviation |
|---|---|---|---|---|
| Dependent Variable | SAP Adoption | 215 | 4.1860 | 5.75535 |
| Covariate | Land Operated | 215 | 857.12 | 2579.599 |
| | Age | 215 | 60.32 | 12.239 |
| | Education level | 215 | 2.6605 | 1.22706 |
| | Income | 215 | 2.7535 | 1.72114 |
| Scale Weight | WEIGHT | 215 | 0.96 | 1.32 |

| Model I: Categorical Variable Information | | | |
|---|---|---|---|
| Variable | Response | N | Percent |
| Crops | Yes | 177 | 82.3% |
| | No | 38 | 17.7% |
| | Total | 215 | 100.0% |
| Veggies | Yes | 21 | 9.8% |
| | No | 194 | 90.2% |
| | Total | 215 | 100.0% |
| Livestock | Yes | 171 | 79.5% |
| | No | 44 | 20.5% |
| | Total | 215 | 100.0% |
| TBP | Yes | 124 | 57.7% |
| | No | 91 | 42.3% |
| | Total | 215 | 100.0% |
| Diverse | Yes | 99 | 46.0% |
| | No | 116 | 54.0% |
| | Total | 215 | 100.0% |
| Inadequate Knowledge | Yes | 23 | 10.7% |
| | No | 192 | 89.3% |
| | Total | 215 | 100.0% |
| Irrigation | Yes | 19 | 8.8% |
| | No | 196 | 91.2% |
| | Total | 215 | 100.0% |
| Work Off farm | Yes | 97 | 45.1% |
| | No | 118 | 54.9% |
| | Total | 215 | 100.0% |
| Sole Proprietorship | Solo | 167 | 77.7% |
| | Others | 48 | 22.3% |
| | Total | 215 | 100.0% |
| Agro tourism Income | Yes | 11 | 5.1% |
| | No | 204 | 94.9% |
| | Total | 215 | 100.0% |

**Table 16: Model II:** NB Regression Model without Income and with Outliers in Land Operated.

| VARIABLES | *B* | STD. ERROR | WALD CHI-SQUARE | SIG. | EXP. (*B*) |
|---|---|---|---|---|---|
| CONSTANT | -0.163 | 0.9014 | 0.033 | 0.857 | 0.850 |
| CROPS*** | 1.097 | 0.2379 | 21.269 | **0.000** | 2.996 |
| VEGGIES* | 0.490 | 0.2792 | 3.076 | 0.079 | 1.632 |
| LIVESTOCK | 0.325 | 0.3350 | 0.938 | 0.333 | 1.383 |
| TBP | 0.194 | 0.2139 | 0.820 | 0.365 | 1.214 |
| DIVERSE*** | 0.770 | 0.2084 | 13.641 | **0.000** | 2.159 |
| INADEQUATE KNOWLEDGE | -0.233 | 0.2701 | 0.744 | 0.389 | 0.792 |
| IRRIGATION** | 0.941 | 0.4230 | 4.954 | **0.026** | 2.564 |
| WORK OFF FARM | 0.040 | 0.2590 | 0.024 | 0.877 | 1.041 |
| SOLE PROPRIETORSHIP | -0.039 | 0.2684 | 0.022 | 0.883 | 0.961 |
| AGRO-TOURISM INCOME** | 1.075 | 0.5040 | 4.552 | **0.033** | 2.931 |
| LAND OPERATED | 0.000 | 0.0003 | 1.055 | 0.304 | 1.000 |
| AGE** | -0.023 | 0.0116 | 3.919 | **0.048** | 0.977 |
| EDUCATION LEVEL*** | 0.302 | 0.0817 | 13.612 | **0.000** | 1.352 |
| INCOME FROM FARMING | -0.163 | 0.9014 | 0.033 | 0.857 | 0.850 |
| *** | 99% Confidence Interval (0.001 Significance Level) | | | | |
| ** | 95% Confidence Interval (0.05 Significance Level) | | | | |
| * | 90% Confidence Interval (0.1 Significance Level) | | | | |
| | | | | | |

| GOODNESS OF FIT | VALUE | DF | VALUE/DF |
|---|---|---|---|
| DEVIANCE | 365.513 | 201 | 1.818 |
| SCALED DEVIANCE | 365.513 | 201 | |
| PEARSON CHI-SQUARED | 387.213 | 201 | 1.926 |
| SCALED PEARSON CHI-SQUARED | 387.213 | 201 | |
| LOG LIKELIHOOD | -407.596 | | |
| AKAIKE'S INFORMATION CRITERION (AIC) | 843.192 | | |
| FINITE SAMPLE CORRECTED AIC (AICC) | 845.292 | | |
| BAYESIAN INFORMATION CRITERION (BIC) | 890.380 | | |
| CONSISTENT AIC (CAIC) | 904.380 | | |
| N | 215 | | |

| Model II: Continuous Variable Information | | N | Mean | Std. Deviation |
|---|---|---|---|---|
| Dependent Variable | SAP Adoption | 215 | 4.1860 | 5.75535 |
| Covariate | Land Operated | 215 | 857.12 | 2579.599 |
| | Age | 215 | 60.32 | 12.239 |
| | Education level | 215 | 2.6605 | 1.22706 |
| | Income | 215 | 2.7535 | 1.72114 |
| Scale Weight | WEIGHT | 215 | 0.96 | 1.32 |

| Model II: Categorical Variable Information | | | |
|---|---|---|---|
| Variable | Response | N | Percent |
| Crops | Yes | 177 | 82.3% |
| | No | 38 | 17.7% |
| | Total | 215 | 100.0% |
| Veggies | Yes | 21 | 9.8% |
| | No | 194 | 90.2% |
| | Total | 215 | 100.0% |
| Livestock | Yes | 171 | 79.5% |
| | No | 44 | 20.5% |
| | Total | 215 | 100.0% |
| TBP | Yes | 124 | 57.7% |
| | No | 91 | 42.3% |
| | Total | 215 | 100.0% |
| Diverse | Yes | 99 | 46.0% |
| | No | 116 | 54.0% |
| | Total | 215 | 100.0% |
| Inadequate Knowledge | Yes | 23 | 10.7% |
| | No | 192 | 89.3% |
| | Total | 215 | 100.0% |
| Irrigation | Yes | 19 | 8.8% |
| | No | 196 | 91.2% |
| | Total | 215 | 100.0% |
| Work Off farm | Yes | 97 | 45.1% |
| | No | 118 | 54.9% |
| | Total | 215 | 100.0% |
| Sole Proprietorship | Solo | 167 | 77.7% |
| | Others | 48 | 22.3% |
| | Total | 215 | 100.0% |
| Agro tourism Income | Yes | 11 | 5.1% |
| | No | 204 | 94.9% |
| | Total | 215 | 100.0% |

**Table 17: Model III**[11]: NB Regression Model without Income and Without Outliers in Land Operated.

| VARIABLES | *B* | STD. ERROR | WALD CHI-SQUARE | SIG. | EXP. (*B*) |
|---|---|---|---|---|---|
| CONSTANT | -0.305 | 0.9138 | 0.111 | 0.739 | 0.737 |
| CROPS*** | 1.067 | 0.2388 | 19.967 | **0.000** | 2.907 |
| VEGGIES** | 0.555 | 0.2816 | 3.892 | **0.049** | 1.743 |
| LIVESTOCK | 0.363 | 0.3386 | 1.147 | 0.284 | 1.437 |
| TBP | 0.181 | 0.2152 | 0.705 | 0.401 | 1.198 |
| DIVERSE*** | 0.778 | 0.2100 | 13.730 | **0.000** | 2.178 |
| INADEQUATE KNOWLEDGE | -0.254 | 0.2706 | 0.882 | 0.348 | 0.776 |
| IRRIGATION** | 0.948 | 0.4261 | 4.954 | **0.026** | 2.582 |
| WORK OFF FARM | 0.109 | 0.2632 | 0.172 | 0.679 | 1.115 |
| SOLE PROPRIETORSHIP | -0.070 | 0.2717 | 0.067 | 0.796 | 0.932 |
| AGRO-TOURISM INCOME* | 0.927 | 0.5187 | 3.193 | 0.074 | 2.526 |
| LAND OPERATED* | 0.001 | 0.0004 | 3.620 | **0.057** | 1.001 |
| AGE* | -0.022 | 0.0118 | 3.611 | **0.057** | 0.978 |
| EDUCATION LEVEL*** | 0.305 | 0.0822 | 13.771 | **0.000** | 1.357 |
| INCOME FROM FARMING | -0.305 | 0.9138 | 0.111 | 0.739 | 0.737 |
| *** | 99% Confidence Interval (0.001 Significance Level) | | | | |
| ** | 95% Confidence Interval (0.05 Significance Level) | | | | |
| * | 90% Confidence Interval (0.1 Significance Level) | | | | |

---

[11] Reference Model.

| GOODNESS OF FIT | VALUE | DF | VALUE/DF |
|---|---|---|---|
| DEVIANCE | 361.122 | 191 | 1.891 |
| SCALED DEVIANCE | 361.122 | 191 | |
| PEARSON CHI-SQUARED | 376.730 | 191 | 1.972 |
| SCALED PEARSON CHI-SQUARED | 376.730 | 191 | |
| LOG LIKELIHOOD | -404.668 | | |
| AKAIKE'S INFORMATION CRITERION (AIC) | 837.337 | | |
| FINITE SAMPLE CORRECTED AIC (AICC) | 839.547 | | |
| BAYESIAN INFORMATION CRITERION (BIC) | 883.859 | | |
| CONSISTENT AIC (CAIC) | 897.859 | | |
| N | 205 | | |

| Model III: Continuous Variable Information | | N | Mean | Std. Deviation |
|---|---|---|---|---|
| Dependent Variable | SAP Adoption | 205 | 4.2000 | 5.79080 |
| Covariate | Land Operated | 205 | 424.53 | 576.883 |
| | Age | 205 | 60.66 | 12.311 |
| | Education level | 205 | 2.6293 | 1.20827 |
| | Income | 205 | 2.6244 | 1.62408 |
| Scale Weight | WEIGHT | 205 | 1.01 | 1.34 |

| Model III: Categorical Variable Information | | | |
|---|---|---|---|
| Variable | Response | N | Percent |
| Crops | Yes | 167 | 81.5% |
| | No | 38 | 18.5% |
| | Total | 205 | 100.0% |
| Veggies | Yes | 21 | 10.2% |
| | No | 184 | 89.8% |
| | Total | 205 | 100.0% |
| Livestock | Yes | 165 | 80.5% |
| | No | 40 | 19.5% |
| | Total | 205 | 100.0% |
| TBP | Yes | 118 | 57.6% |
| | No | 87 | 42.4% |
| | Total | 205 | 100.0% |
| Diverse | Yes | 93 | 45.4% |
| | No | 112 | 54.6% |
| | Total | 205 | 100.0% |
| Inadequate Knowledge | Yes | 23 | 11.2% |
| | No | 182 | 88.8% |
| | Total | 205 | 100.0% |
| Irrigation | Yes | 17 | 8.3% |
| | No | 188 | 91.7% |
| | Total | 205 | 100.0% |
| Work Off farm | Yes | 96 | 46.8% |
| | No | 109 | 53.2% |
| | Total | 205 | 100.0% |
| Sole Proprietorship | Solo | 165 | 80.5% |
| | Others | 40 | 19.5% |
| | Total | 205 | 100.0% |
| Agro tourism Income | Yes | 10 | 4.9% |
| | No | 195 | 95.1% |
| | Total | 205 | 100.0% |

**Table 18: Model IV:** NB Regression Model with Income and without Outliers in Acres Operated.

| VARIABLES | *B* | STD. ERROR | WALD CHI-SQUARE | SIG. | EXP. (*B*) |
|---|---|---|---|---|---|
| CONSTANT | -0.282 | 0.9217 | 0.093 | 0.760 | 0.755 |
| CROPS*** | 1.078 | 0.2457 | 19.247 | **0.000** | 2.939 |
| VEGGIES* | 0.547 | 0.2856 | 3.665 | **0.056** | 1.728 |
| LIVESTOCK | 0.365 | 0.3384 | 1.162 | 0.281 | 1.440 |
| TBP | 0.185 | 0.2163 | 0.732 | 0.392 | 1.203 |
| DIVERSE*** | 0.782 | 0.2110 | 13.741 | **0.000** | 2.186 |
| INADEQUATE KNOWLEDGE | -0.252 | 0.2710 | 0.867 | 0.352 | 0.777 |
| IRRIGATION** | 0.947 | 0.4262 | 4.933 | **0.026** | 2.577 |
| WORK OFF FARM | 0.100 | 0.2672 | 0.141 | 0.707 | 1.106 |
| SOLE PROPRIETORSHIP | -0.071 | 0.2716 | 0.068 | 0.795 | 0.932 |
| AGRO-TOURISM INCOME | 0.915 | 0.5222 | 3.069 | **0.080** | 2.496 |
| LAND OPERATED* | 0.001 | 0.0005 | 2.921 | **0.087** | 1.001 |
| AGE* | -0.023 | 0.0119 | 3.648 | **0.056** | 0.978 |
| EDUCATION LEVEL*** | 0.309 | 0.0851 | 13.208 | **0.000** | 1.362 |
| INCOME FROM FARMING | -0.020 | 0.1071 | 0.036 | 0.850 | 0.980 |
| *** | 99% Confidence Interval (0.001 Significance Level) | | | | |
| ** | 95% Confidence Interval (0.05 Significance Level) | | | | |
| * | 90% Confidence Interval (0.1 Significance Level) | | | | |

| Model IV: Continuous Variable Information | | N | Mean | Std. Deviation |
|---|---|---|---|---|
| Dependent Variable | SAP Adoption | 205 | 4.2000 | 5.79080 |
| Covariate | Land Operated | 205 | 424.53 | 576.883 |
| | Age | 205 | 60.66 | 12.311 |
| | Education level | 205 | 2.6293 | 1.20827 |
| | Income | 205 | 2.6244 | 1.62408 |
| Scale Weight | WEIGHT | 205 | 1.01 | 1.34 |

| GOODNESS OF FIT | VALUE | DF | VALUE/DF |
|---|---|---|---|
| DEVIANCE | 361.086 | 190 | 1.900 |
| SCALED DEVIANCE | 361.086 | 190 | |
| PEARSON CHI-SQUARED | 377.713 | 190 | 1.988 |
| SCALED PEARSON CHI-SQUARED | 377.713 | 190 | |
| LOG LIKELIHOOD | -404.650 | | |
| AKAIKE'S INFORMATION CRITERION (AIC) | 839.301 | | |
| FINITE SAMPLE CORRECTED AIC (AICC) | 841.841 | | |
| BAYESIAN INFORMATION CRITERION (BIC) | 889.146 | | |
| CONSISTENT AIC (CAIC) | 904.146 | | |
| N | 205 | | |

| Model IV: Categorical Variable Information | | | |
|---|---|---|---|
| **Variable** | **Response** | **N** | **Percent** |
| **Crops** | **Yes** | 167 | 81.5% |
| | **No** | 38 | 18.5% |
| | **Total** | 205 | 100.0% |
| **Veggies** | **Yes** | 21 | 10.2% |
| | **No** | 184 | 89.8% |
| | **Total** | 205 | 100.0% |
| **Livestock** | **Yes** | 165 | 80.5% |
| | **No** | 40 | 19.5% |
| | **Total** | 205 | 100.0% |
| **TBP** | **Yes** | 118 | 57.6% |
| | **No** | 87 | 42.4% |
| | **Total** | 205 | 100.0% |
| **Diverse** | **Yes** | 93 | 45.4% |
| | **No** | 112 | 54.6% |
| | **Total** | 205 | 100.0% |
| **Inadequate Knowledge** | **Yes** | 23 | 11.2% |
| | **No** | 182 | 88.8% |
| | **Total** | 205 | 100.0% |
| **Irrigation** | **Yes** | 17 | 8.3% |
| | **No** | 188 | 91.7% |
| | **Total** | 205 | 100.0% |
| **Work Off farm** | **Yes** | 96 | 46.8% |
| | **No** | 109 | 53.2% |
| | **Total** | 205 | 100.0% |
| **Sole Proprietorship** | **Solo** | 165 | 80.5% |
| | **Others** | 40 | 19.5% |
| | **Total** | 205 | 100.0% |
| **Agro tourism Income** | **Yes** | 10 | 4.9% |
| | **No** | 195 | 95.1% |
| | **Total** | 205 | 100.0% |

**Appendix G: Survey Tools and Coding** (Only relevant are included.)

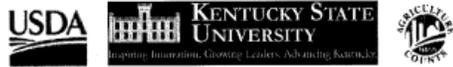

2015 FARM SUSTAINABILITY, DIVERSIFICATION AND EFFICIENCY SURVEY
KENTUCKY STATE UNIVERSITY

**This questionnaire must be received no later than September 11, 2015.**

After data collection, the National Agricultural Statistics Service (NASS) processes the data independent of names and addresses. Under Title 7 of the U.S. Code and CIPSEA (Public Law 107-347), facts about your operation are kept confidential and used only for statistical purposes in combination with similar reports from other producers. Please have the farm's decision makers answer the questions. Thanks for the time.

ID: T_0001 ______

Name: T_0002 ______

Address (the majority of your farm): T_0003 ______ County: T_0004 ______

City: T_0005 ___ State: T_0006 ___ ZIP Code: T_0007 ______

(Check one) **Submitted via:**

C_0012 The above information is correct

C_0013 The above information is incorrect

| C_0008 | Mail |
|---|---|
| C_0009 | Phone |
| C_0010 | Other: T_0011 |

Coded to give a character output.
Mode = 'phone'.

---

**SECTION 3: Sustainability**

Definition: Sustainable agriculture is the production of food, fiber, or other plant or animal products using farming techniques that protect the environment, public health, human communities, and animal welfare. This form of agriculture enables us to produce healthy food without compromising future generations' ability to do the same.

1. Overall, how would you rate your farming practices as sustainable? (Check one value for each choice, where 1 is not sustainable and 5 is very sustainable)

| Indicators of Sustainability | 1= Not Sustainable 5= Very Sustainable | | | | |
|---|---|---|---|---|---|
|  | 1 | 2 | 3 | 4 | 5 |
| Local ecosystem sustainability (e.g., water, soil, air, bio-diversity, and critical species) | C_0 150 | C_0 151 | C_ 01 52 | C_ 01 53 | C_ 01 54 |
| Global ecosystem sustainability (e.g., energy use, climate change) | C_0 155 | C_0 156 | C_ 01 57 | C_ 01 58 | C_ 01 59 |
| Social sustainability (e.g., quality of life, labor, economics, community) | C_0 160 | C_0 161 | C_ 01 62 | C_ 01 63 | C_ 01 64 |
| Social acceptability (e.g., community regulations) | C_0 165 | C_0 166 | C_ 01 67 | C_ 01 68 | C_ 01 69 |
| Food system sustainability (e.g., food quality, nutrition, affordability and access) | C_0 170 | C_0 171 | C_ 01 72 | C_ 01 73 | C_ 01 74 |

a
b
c
d
e

---

**Section 1: Farm details**

1. How many acres of land did you operate in 2014?

| Owned | N_0014 acres |
|---|---|
| Rented from | N_0015 acres |
| Total acres operated | n_0999 acre |

2a. Do you grow crops?C_0016Yes (go to part 2b)

Var = crops      C_0017 No (go to question 3a)

2b. Check and estimate how many acres of crops were grown in 2014. (leave blank for 0)

if filled out this becomes "1"

| Corn | N_0018Acres | Sweet sorghum | N_0019Acres |
|---|---|---|---|
| Soybean | N_0020Acres | Energy Cane | N_0021Acres |
| Tobacco | N_0022Acres | Switch grass | N_0023Acre |
| Sorghum | N_0024Acres | Miscanthus | N_0025Acres |
| Hay crop | N_0026Acres | Other1: T_0027 | N_0028Acres |
| Barley | N_0029Acres | Other2: T_0030 | N_0031Acres |

3a. Do you grow fruits and/or vegetables?C_0032Yes (go to part 3b)

Var = veggies      C_0033 No (go to question 4)

3b. Check and estimate how many acres of fruits/vegetables were grown in 2014, excluding home gardens. (leave blank for 0)

| Vegetables in tunnel/greenhouse/hydroponics | N_0034Acres | Fruits and Nuts | N_0035Acre |
|---|---|---|---|
| Vegetables grown in farm | N_0036Acres | Nursery (trees) | N_0037Acre |
| Melons | N_0038Acres | Berries | N_0039Acre |
| Potatoes | N_0040Acres | Other1: T_0041 | N_0042Acre |
| Sweet Potatoes | N_0043Acres | Other2: T_0044 | N_0045Acre |

3c. Since 2010, have you changed the cropping pattern (crop mix)?

C_0046YES (continue)

C_0047NO (go to question 4a)    Var = Sec1Q3c   where yes=1, no=0

7a. Have you adopted/participated in any of these practices in farming? (Check one for each practice)

| | Sustainable practices | Yes, Adopted | Not Adopted | Planning to adopt | Not Applicable |
|---|---|---|---|---|---|
| a. | Increasing biological diversity (planting perennial crops along with annual crops, fragment land with forest, riparian areas and agriculture landscapes to foster multi species existence ) | C_02 06 | C_02 07 | C_02 08 | C_02 09 |
| b. | Reduced chemical pesticide usage | C_02 10 | C_02 11 | C_02 12 | C_02 13 |
| c. | Reduced chemical fertilizer usage | C_02 14 | C_02 15 | C_02 16 | C_02 17 |
| d. | Cover cropping or green manure | C_02 18 | C_02 19 | C_02 20 | C_02 21 |
| e. | Crop rotation | C_02 22 | C_02 23 | C_02 24 | C_02 25 |
| f. | Biological pest control (using natural enemies of pests) | C_02 26 | C_02 27 | C_02 28 | C_02 29 |
| g. | Cultural pest control (modification of cultivation practices such as using trap crops, sanitation, planting and harvesting date variation) | C_02 30 | C_02 31 | C_02 32 | C_02 33 |
| h. | Manure distribution as fertilizer | C_02 34 | C_02 35 | C_02 36 | C_02 37 |
| i. | Alley cropping (crops growing in alleys formed between trees or shrubs) | C_0 238 | C_0 239 | C_0 240 | C_0 241 |
| j. | Integration of crop and livestock production systems | C_02 42 | C_02 43 | C_02 44 | C_02 45 |
| k. | Variety mixtures of single crops | C_02 46 | C_02 47 | C_02 48 | C_02 49 |
| l. | Growing native or "local" crops | C_02 50 | C_02 51 | C_02 52 | C_02 53 |
| m. | Poly-culture farming (more than one crop grown in a field at the same time) | C_02 54 | C_02 55 | C_02 56 | C_02 57 |
| n. | Crop and livestock diversification | C_02 58 | C_02 59 | C_02 60 | C_02 61 |
| o. | Controlled grazing (rotational, intensive, etc.) | C_02 62 | C_02 63 | C_02 64 | C_02 65 |
| p. | Use of animals to control or eliminate brush for land reclamation | C_02 66 | C_02 67 | C_02 68 | C_02 69 |

Sec3Qa,a
b
c

| | | | | | |
|---|---|---|---|---|---|
| q. | Multiple-species grazing (grazing goat or sheep with cattle) | C_02 70 | C_02 71 | C_02 72 | C_02 73 |
| r. | Forest stewardship (long-term stewardship of nonindustrial private forest lands by more actively managing their forest and related resources) | C_02 74 | C_02 75 | C_02 76 | C_02 77 |
| s. | Farm machinery adaptations to promote erosion control | C_02 78 | C_02 79 | C_02 80 | C_02 81 |
| t. | Composting | C_02 82 | C_02 83 | C_02 84 | C_02 85 |
| u. | Ridge till (The soil is left undisturbed from harvest to planting except for strips up to 1/3 of the row width.) | C_02 86 | C_02 87 | C_02 88 | C_02 89 |
| v. | Windbreaks and/or shelterbelts | C_02 90 | C_02 91 | C_02 92 | C_02 93 |
| w. | Expert computer systems for farm management or precision agriculture | C_02 94 | C_02 95 | C_02 96 | C_02 97 |
| x. | Fallow management systems | C_02 98 | C_02 99 | C_03 00 | C_03 01 |
| y. | Mulching | C_03 02 | C_03 03 | C_03 04 | C_03 05 |
| z. | Sprayer calibration and application accuracy | C_03 06 | C_03 07 | C_03 08 | C_03 09 |
| aa. | Improved water management techniques (drainage and irrigation) | C_03 10 | C_03 11 | C_03 12 | C_03 13 |
| bb. | Land farming to reduce erosion (terracing, contour planting, grade stabilization, etc.) | C_03 14 | C_03 15 | C_03 16 | C_03 17 |
| cc. | Integrated pest management techniques | C_03 18 | C_03 19 | C_03 20 | C_03 21 |
| dd. | Conservation tillage in cropping systems | C_03 22 | C_03 23 | C_03 24 | C_03 25 |
| ee. | Reforestation | C_03 26 | C_03 27 | C_03 28 | C_03 29 |

7b. If you have not adopted sustainable practices provided in part a above, why not? (*Check all that apply; leave blank if answered 7a*).

| | |
|---|---|
| C_0330 | Inadequate knowledge |
| C_0331 | Perceived difficulty of implementation |
| C_0332 | Pressure to increase crop/livestock productivity |
| C_0333 | Lack of adequate markets for alternative products |
| C_0334 | Lack of consumer acceptance for alternative products |
| C_0335 | Negative attitude about these technologies/practices |
| C_0336 | Lack of appropriate technology |
| C_0337 | Happy with what I am doing currently |

**SECTION 6: Demographic questions**

1. How old were you on your last birthday? N_0610 **years old**

2. What is your highest level of formal education (*check one*)?

*Sec6Q2-a* *Sec6Q2-b*

| | You | | Your Spouse |
|---|---|---|---|
| 1 | C_0611 | Below high school. | C_0612 | Below high school. |
| 2 | C_0613 | High school degree. | C_0614 | High school degree. |
| 3 | C_0615 | Attended some college. | C_0616 | Attended some college. |
| 4 | C_0617 | College degree. | C_0618 | College degree. |
| 5 | C_0619 | Professional or graduate school degree. | C_0620 | Professional or graduate school degree. |

0 = multi

4. How long have you been farming? N_0621 **years**

5a. How long have you been making farming decisions? N_0622 **years**

5b. What is your gross income from farming in 2014 (*check one*)?

*Sec6Q5b*

| | | |
|---|---|---|
| 1 | C_0623 | less than $10,000 |
| 2 | C_0624 | $10,000-less than $49,999 |
| 3 | C_0625 | $50,000-less than $99,999 |
| 4 | C_0626 | $100,000- less than $249,999 |
| 5 | C_0627 | $250,000-less than $499,999 |
| 6 | C_0628 | more than $500,000 |

6a. What was the amount received in 2014 for all Federal/State agricultural program payments (*check one*)?

*Sec6Q6a*

| | | |
|---|---|---|
| 0 | C_0629 | None |
| 1 | C_0630 | less than $1,000 |
| 2 | C_0631 | $1,000-less than $4,999 |
| 3 | C_0632 | $5,000-less than $9,999 |
| 4 | C_0633 | $10,000- less than $24,999 |
| 5 | C_0634 | $25,000-less than $49,999 |
| 6 | C_0635 | more than $50,000 |

6b. What percentage of farm income from the previous question is from adopting conservation reserve (CRP), wetland reserve (WRP), farmable wetlands (FWP) or conservation reserve enhancement (CREP) programs? N_0636 **%**

7. Do you get income from Agri-tourism or recreational services (e.g. farm tours, hunting rights, leasing)?
C_0637Yes C_0638No

*Sec6Q7*

8. How far are you from the closest town? N_0639 **miles**

9a. How far are you from the closest seed/fertilizer dealer? N_0640 **miles**

9b. Does the closest seed/fertilizer dealer sell organic seed?
C_0641Yes C_0642No C_0643Don't know

*Sec6Q9b*

10. How far are you from closest farm equipment dealer? N_0644 **miles**

11a. Do you or your spouse work off farm?
C_0645Yes (*go to 11b*)
C_0646No (*go to question 12a*)

*Sec6Q11a*

11b. How many hours per week do you work off farm? N_0647 **hours per week**

11c. How many hours per week does your spouse work off farm? N_0648 **hours per week**

11d. What is the primary reason for working off farm (*Check all that apply*):

| | |
|---|---|
| C_0649 | Health Benefits |
| C_0650 | Supplemental income |
| C_0651 | Retirement benefits |
| C_0652 | Other (specify): T_0653 |

12a. Are you of Spanish, Hispanic, or Latino origin or background, such as Mexican, Cuban, or Puerto Rican, regardless of race?
C_0654Yes C_0655No

*Sec6Q12a*

12b. What is your race? (*check all that apply*)

| | |
|---|---|
| C_0656 | Black or African American |
| C_0657 | White |
| C_0658 | Asian |
| C_0659 | Native American or Alaska Native |
| C_0660 | Native Hawaiian or other Pacific Islander |
| C_0661 | Other (specify): T_0662 |

13. What is your type of farm operation?

| | |
|---|---|
| C_0663 | Sole proprietorship |
| C_0664 | Partnership |
| C_0665 | Family held corporation |
| C_0666 | Corporation (not family held) |
| C_0667 | Cooperative |
| C_0668 | Trust or estate |
| C_0669 | Institutional |
| C_0670 | Other (specify): T_0671 |

14. Did you participate in the tobacco buy-out program of 2004? C_0672Yes C_0673No C_0674NA

*Sec6Q14*

15. In your opinion, what are alternatives to tobacco production (*Check all that apply*)

| | |
|---|---|
| C_0675 | Hemp |
| C_0676 | Blueberries |
| C_0677 | Commercial poultry |
| C_0678 | Free range poultry |
| C_0679 | Aquaculture |
| C_0680 | Other crops/livestock enterprises (specify): T_0681 |

**Appendix H: Focus Group Discussion Pictures.**

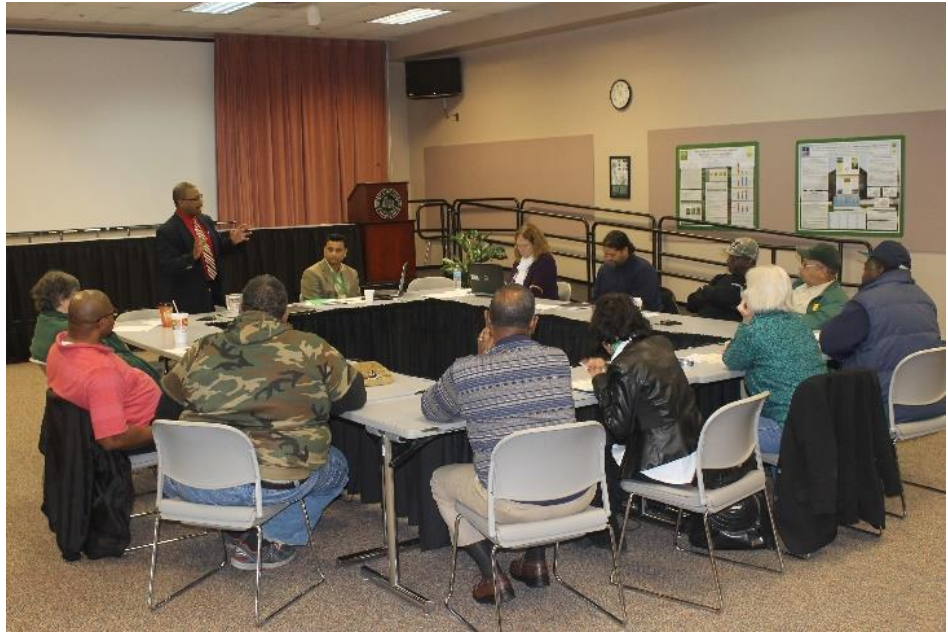

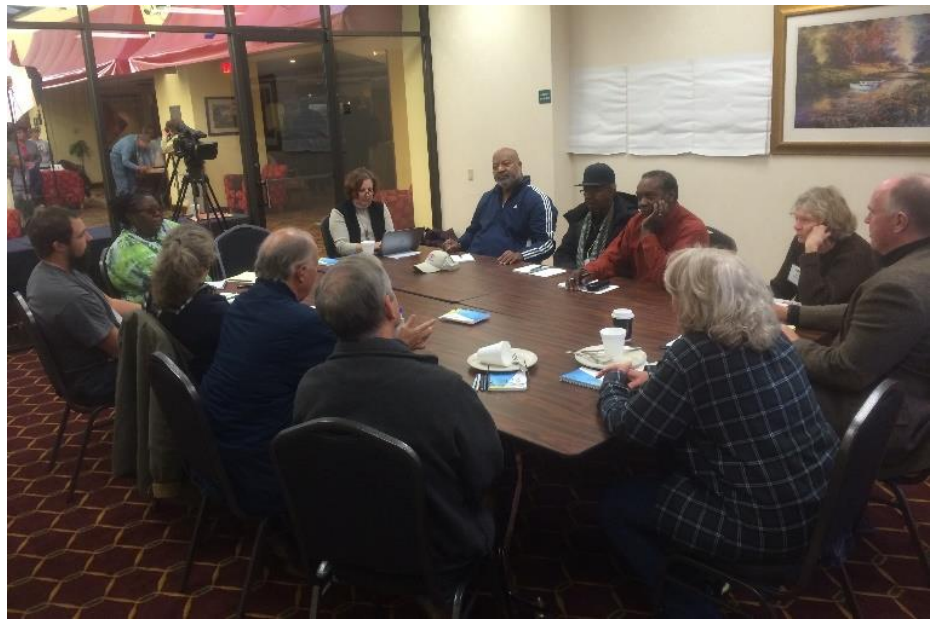

Figure 5: Two focus group discussions conducted in Kentucky State University and Field to get insight of agriculture before the research and designing survey.

conditioning. *Cercetaria Agronomice in Moldova,XLIII*(3 (143)), 91-1000.